 \documentclass[twocolumn,a4paper,traditabstract]{aa}
\usepackage{amsmath}
\usepackage{graphicx,txfonts,natbib,verbatim}
\usepackage{color}
\bibpunct{(}{)}{;}{a}{}{,}

\begin{document}

\title{Plasma radio emission from inhomogeneous collisional plasma of a flaring loop}

\author{H.~Ratcliffe\inst{\ref{inst1},\ref{inst2}}\and E.P.~Kontar\inst{\ref{inst2}}}

\institute{Centre for Space, Fusion and Astrophysics, University of Warwick, CV4 7AL \label{inst1} \and School of Physics \& Astronomy, University of Glasgow, G12 8QQ, United Kingdom \label{inst2}}

\offprints{H. Ratcliffe \email{h.ratcliffe@warwick.ac.uk}}


\date{Received 11/07/13 ; Accepted 05/12/13}

\abstract{The evolution of a solar flare accelerated non-thermal electron population and associated plasma
emission is considered in collisional inhomogeneous plasma. Non-thermal electrons collisionally evolve
to become unstable and generate Langmuir waves, which may lead to intense radio emission.
We self-consistently simulated the collisional relaxation of electrons, wave-particle interactions,
and non-linear Langmuir wave evolution in plasma with density fluctuations.
Additionally, we simulated the scattering, decay, and coalescence of the Langmuir
waves which produce radio emission at the fundamental or the harmonic of the plasma frequency,
using an angle-averaged emission model. Long-wavelength density fluctuations, such as are observed in the corona,
are seen to strongly suppress the levels of radio emission, meaning that a high level of Langmuir waves
can be present without visible radio emission. Additionally, in homogeneous plasma, the emission shows time
and frequency variations that could be smoothed out by density inhomogeneities.}

\keywords{Sun:particle emission, Sun: radio emission}
\titlerunning{Plasma radio emission in inhomogeneous plasma}
\authorrunning{Ratcliffe and Kontar}

\maketitle
\section{Introduction}

During solar flares, efficient acceleration of electrons in coronal loops is often observed.
Such non-thermal electrons produce several radiation signatures, in particular hard
X-rays (HXR), and radio emissions. These signatures are a valuable diagnostic
for the properties of non-thermal electron populations.
HXR emission is produced by collisional bremsstrahlung in the dense plasma of loop footpoints,
and is particularly useful because the coronal plasma is optically thin to these wavelengths,
and the cross-section for emission well known. Gyrosynchrotron emission (gyroemission from mildly relativistic particles)
can produce strong continuum radio emission at GHz frequencies in regions of high magnetic field,
while coherent emission mechanisms can produce intense radio bursts in some circumstances.

In particular, as noted by \citet{1984ApJ...279..882E}, \citet{1987ApJ...321..721H} fast electrons
in dense plasma can produce Langmuir wave turbulence {because of collisional evolution, and this can lead to intense plasma
radio emission due to the large number of non-thermal electrons in flares. {This collisional relaxation} is the mechanism we consider here.}
For emission at the harmonic of the plasma
frequency, resulting from the coalescence of two counter propagating Langmuir waves,
a simple analytical estimate shows that this process should saturate at a brightness temperature equal
to the brightness temperature of the Langmuir waves, therefore often reaching several orders of magnitude
over the thermal level and thus easily visible.

On the other hand, Langmuir waves are known to be strongly affected by fluctuations in plasma density. Spreading of Langmuir waves in angle because of elastic scattering by density fluctuations was considered by e.g \citet{1976JPSJ...41.1757NF} and \citet{1985SoPh...96..181M} and found to suppress the beam-plasma interaction. A similar result for beam-aligned density fluctuations was found in our previous work \citep{RBK}.
Moreover, Langmuir waves generated in non-uniform plasma can be shifted to lower wave numbers (higher phase velocities)
and re-absorbed, leading to efficient acceleration of the tail of energetic electrons. A relatively high level of Langmuir waves can
strongly affect the electron distribution above $\sim 20$~keV and lead to substantial over-estimation of electron number and energy
in flares when simple collisional relaxation is assumed \citep{KRB}. This self-acceleration of the electron tail
is particularly efficient when the timescale of Langmuir wave refraction is similar to the timescale of Langmuir wave
generation \citep{RBK} and is evident in 3D PIC simulations in magnetised plasma \citep{2012A&A...544A.148K}.

{Thus it appears that density fluctuations can suppress the Langmuir wave generation from an electron population, and therefore allow for the existence of fast electrons without plasma radio emission.} However, it still remains unclear whether a high level of Langmuir turbulence necessitates strong plasma emission in collisional
flaring plasma, {since a density gradient acting on Langmuir waves has already been found by \citet{2002PhRvE..65f6408K} to suppress Langmuir wave decay, potentially allowing the production of Langmuir waves without their conversion into radio emission.} {Addressing this problem requires detailed numerical simulations, because although} the processes involved in the emission are qualitatively known \citep[e.g.][]{1958SvA.....2..653G,1964NASSP..50..357S,1970SvA....14...47Z},
an analytical treatment is not possible. Such numerical simulations are
often employed for calculations applied to type III bursts \citep[e.g.][]{2008JGRA..11306104L,2011PhPl...18e2903T}.

Here we consider the plasma emission from accelerated electrons in a dense coronal loop. We
self-consistently solve the problem of relaxation of a non-thermal electron population taking
Langmuir wave excitation, absorption, and weakly-nonlinear evolution into account.
We show that plasma emission will be produced at both the fundamental and the harmonic of the plasma frequency,
reaching brightness temperatures up to several orders of magnitude above thermal, although due to the effects
of escape through the surrounding plasma, only the harmonic component can be observed. {The emission has an intrinsic bandwidth of the order of $5\%$ of the plasma frequency due to its wavenumber spread} {and the dispersive nature of electromagnetic waves in plasma.}
In homogeneous density plasma, this emission shows significant time and frequency variations,
while the presence of density fluctuations in the background plasma is seen to both smooth out these variations
and reduce the brightness of the emission. A relatively modest level of density fluctuations is able to completely
suppress the emission, despite the presence of a high level of Langmuir waves, while simultaneously producing
electron self-acceleration and thus increasing the number of electrons at high velocities.

\section{A model for electron and Langmuir wave evolution, and plasma radio emission}

\begin{figure}
\centering
\includegraphics[width=0.49\textwidth]{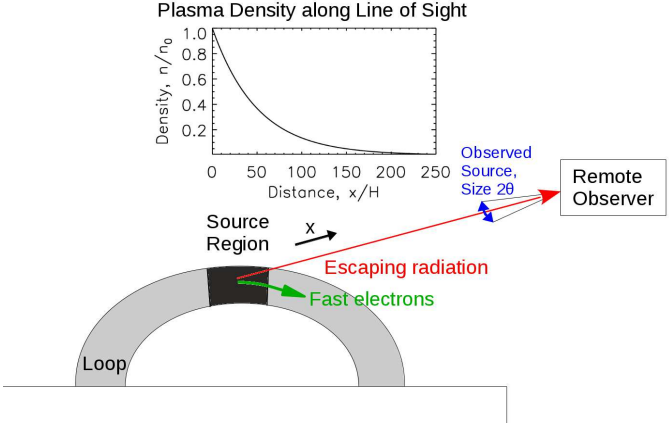}
\caption{Geometry of a model coronal loop. Within the loop the plasma density is approximately $n_e$, with superimposed weak random density fluctuations. A fast electron population is produced in the source region, {shaded dark grey}, which relaxes collisionally, producing Langmuir waves and thus radio emission. {Radio emission arises from the source region of linear size $d\simeq 10^9$~cm {measured perpendicular to the loop axis}, of the order of the typical cross-sectional size of a coronal loop.} An observer at 1AU sees a source with angular radius $\theta$. The surrounding plasma has an exponential density profile of scale height $H$.}
\label{fig:lpGeom}
\end{figure}

We consider a dense coronal loop embedded in the coronal plasma, as shown in Figure \ref{fig:lpGeom}.
The electron density is the combination of a constant density background plus small magnitude, long wavelength random
density fluctuations. The evolution of a non-thermal electron population is considered in the collisional
region, {with the electron distribution, and all wave distributions, assumed not to depend on position within the source volume}. The electron beam-plasma system is self-consistently
treated taking into account the effects of non-uniform plasma density, plasma collisionality, non-linear wave-wave
processes and the production of electromagnetic radio emission near the plasma frequency and its harmonic.

All of these processes are interrelated, and therefore simulated simultaneously. {The equations describing the evolution of the electron distribution are given in Section \ref{sec:Els}, and those for Langmuir waves and ion-sound waves in Sections \ref{sec:Lang} and \ref{sec:Ion} respectively. The basic equations for electromagnetic emission are in Section \ref{sec:EM}, with the emission at the fundamental of the plasma frequency treated in Section \ref{sec:Fund}, and the harmonic in Section \ref{sec:Harm}. Finally, Section \ref{sec:Prop} addresses the considerations of propagation and absorption of electromagnetic emission.} We note that the dynamics of electrons, Langmuir waves and ion-sound waves produced from Langmuir wave decay are all approximately one-dimensional. On the other hand, the equations for Langmuir wave conversion into electromagnetic emission are not, and are treated
here using an angle-averaged approximation, described briefly below.

\subsection{Non-thermal electron evolution}\label{sec:Els}

We describe the electrons using their distribution function $f({\mathrm v},t)$ [cm$^{-3}$/(cm s$^{-1}$)],
which is normalised so that
\begin{equation}\int_{-\infty}^{\infty} f({\mathrm v},t)\, {\mathrm d}{\mathrm v}=n_e +n_b,
\end{equation} {where $n_e$ the density of the background plasma, and $n_b$ that of non-thermal electrons,}
and the Langmuir waves by their spectral energy density as a function of wavenumber $k$, $W(k,t)$ [erg cm$^{-2}$],
where
\begin{equation}\int_{-k_{De}}^{k_{De}} W(k,t)\, {\mathrm d} k=E_L,
\end{equation}
is the total energy density of the waves in erg cm$^{-3}$ and $k_{De}=\omega_{pe}/{\mathrm v}_{Te}$ {with
${\mathrm v}_{Te}=\sqrt{k_B T_e/m_e}$ the electron thermal velocity and $\omega_{pe}=\sqrt{4\pi n_e e^2/m_e}$
the plasma frequency, and $m_e, e$ the mass and charge of an electron respectively.}

The equation for the evolution of the electron distribution is {based on that in \citep{drummond1964nucl, vedenov1962quasi} with additional terms describing the effects of collisions. We have
 \begin{align}\label{eqn:4ql1}
\frac{\partial f( {\mathrm v},t)}{\partial t}= \frac{4\pi^2
e^2}{m_e^2}\frac{\partial}{\partial {\mathrm v}}\left( \frac{W(k,t)}{{\mathrm
v}}\frac{\partial f( {\mathrm v},t)}{\partial {\mathrm v}}\right) + \notag \\
\Gamma\frac{\partial}{\partial \mathrm v}\left(\frac{f( {\mathrm v},t)}{\mathrm
v^2}+\frac{\mathrm v_{Te}^2}{\mathrm v^3}\frac{\partial f( {\mathrm v},t)}{\partial \mathrm
v}\right),
\end{align}
with
$\Gamma = 4 \pi e^4 n_e \ln \Lambda/m_e^2$ and $\ln\Lambda$ the Coulomb logarithm,
approximately 20 in the solar corona. }

The first term on the RHS of this equation describes emission and absorption of Langmuir waves
by the electrons \citep{drummond1964nucl, vedenov1962quasi,1963JNuE....5..169V}. This is a resonant interaction between electrons
and Langmuir waves, $\omega_{pe}=k{\mathrm v}$ \footnote{{More precisely, $\omega_{L}=k{\mathrm v}$ with $\omega_L\simeq \omega_{pe} +3k^2 {\mathrm v}_{Te}^2/(2\omega_{pe})$ the Langmuir wave frequency should be considered, but
since $k/k_{De} <<1$  for the excited Langmuir waves, $\omega_{L}\approx \omega_{pe}$  is good approximation
\citep{1963JNuE....5..169V}.}}, i.e. the Langmuir wave phase velocity must equal the electron velocity. {The second and third terms on the RHS describe
collisional interaction of electrons with Maxwellian plasma \citep[e.g.][]{1981phki.book.....L}: the second is the total energy loss
and the third term accounts for the finite temperature of the background electrons. When the level of plasma waves is near
thermal, the first term on the RHS becomes negligible and the stationary ($\partial f/\partial t=0$) Maxwellian distribution is formed by collisions,
i.e. by the second and the third terms.} {This collisional operator is approximately proportional to} $1/{\mathrm v}^3$ {and so the slower electrons will lose energy more rapidly than the faster ones, leading to a reverse slope in velocity space, which will lead to generation of Langmuir waves. Langmuir wave generation will tend to flatten this reverse slope, while collisions continue to restore it, leading to a plateau in the electron distribution with slowly decreasing height as the electrons thermalise.}

\subsection{Langmuir wave evolution}\label{sec:Lang}

The electron beam generates Langmuir waves (denoted L) with wavenumbers parallel to the direction of propagation.
These may then be backscattered by two processes $L\rightleftarrows L' + s$
involving ion-sound waves (denoted s) and $L + i\rightleftarrows L' + i'$ where $i,i'$ denote initial and final {states of a} plasma ion.
The relative importance of these two processes depends strongly on the ratio of electron and ion temperatures, $T_e/T_i$, which can vary between $\sim 2$ and 0.1 in the corona and solar wind \citep[e.g.][]{1979JGR....84.2029G,1998JGR...103.9553N}.
Generally, ion scattering is important in plasma with $T_i \ge T_e$, where ion-sound waves are very strongly
damped, and the decay involving ion-sound waves
when the ion temperature is lower \citep[e.g.][]{2000PhPl....7.4901C}.

The governing equation for the spectral energy density of Langmuir waves is  \citep{drummond1964nucl, vedenov1962quasi,1970SvA....14...47Z}
\begin{align}\label{eqn:4ql2}&
\frac{\partial W(k,t)}{\partial t} =\frac{\omega_{pe}^3 m_e}{4\pi n_e}{\mathrm v}\ln\left(\frac{\mathrm v}{\mathrm v_{Te}}\right)f({\mathrm v},t)
+\frac{\pi\omega_{pe}^3}{n_ek^2}W(k,t)\frac{\partial f( {\mathrm v},t)}{\partial {\mathrm
v}}\notag \\ &-\gamma_{c}W(k,t)+
\frac{\partial}{\partial
k}\left(D(k)\frac{\partial W(k,t)}{\partial k}\right)\notag \\ & + {\rm St}_{decay}(W(k,t))+{\rm St}_{ion}(W(k,t)),
\end{align}
where $\gamma_{c}=\frac{1}{3}\sqrt{(2/\pi)}\; \Gamma /{\mathrm v}_{Te}^3\simeq {\Gamma}/(4 {\mathrm v}_{Te}^3)$, and $k=\omega_{pe}/{\mathrm v}$.
{The first and second terms on the RHS are the spontaneous and stimulated emission of Langmuir waves
by electrons. These terms are derived in the weak-turbulence limit,
\begin{equation}
E_L/(n_ek_B T_e)\ll (k/k_{De})^2,
\end{equation}
so that the energy density of Langmuir waves is much less than the energy density of surrounding plasma.}

Collisional damping of Langmuir waves \citep[e.g.][]{1981phki.book.....L,1980gbs..bookR....M}
is given by the third term on the RHS in Equation (\ref{eqn:4ql2}), while the fourth describes Langmuir wave
diffusion in wavenumber space due to long-wavelength plasma density fluctuations.
For a plasma density which is given by $n(x,t)=n_e(1+\tilde{n}(x,t))$, the sum of a background and small fluctuations
$\tilde{n} \ll 1$ with length scales far larger than the Debye length, the diffusion coefficient is \citep{RBK}
\begin{equation}\label{eqn:diffcoeffGauss}
D(k)= \omega_{pe}^2{\pi}^{3/2}\frac{q_0}{{{\mathrm v}}_0}\langle\tilde{n}^2\rangle \left(1+\frac{{{\mathrm v}}_g(k)^2}{{ {\mathrm v}}_0^2}\right)^{-3/2},
\end{equation}
for random density fluctuations with characteristic velocity ${\mathrm v}_0$, wavenumber $q_0$
and RMS (root mean square) fluctuation level $\sqrt{\langle \tilde{n}^2\rangle}$, and ${\mathrm v}_g= 3 {\mathrm v}_{Te}^2 k/\omega_{pe}$ the group velocity of Langmuir waves.

The two terms labelled ${\mathrm St}$ describe the scattering and decay of Langmuir waves by the processes $L + i\rightleftarrows L' + i'$ and $L\rightleftarrows L' + s$ where $i,i'$ are an initial
and scattered plasma ion and $s$ denotes an ion-sound wave.
Both of these produce back-scattered (negative wavenumber) Langmuir waves. {The general expressions describing these processes are standard \citep[e.g.][]{1980MelroseBothVols, 1995lnlp.book.....T}, and have been rewritten here under the assumption that fast-electron generated Langmuir waves propagate approximately parallel to the generating electrons and therefore the ambient magnetic field. Hereafter, we omit the explicit time dependence of the spectral energy densities for clarity of notation.}

{Langmuir wave scattering by ions is described by}
\begin{align}\label{eqn:LIonA}
{\rm St}_{ion}(W(k))=\int dk_{L'} \frac{\alpha_{ion}}{|k-k_{L'}|} \exp{\left(-\frac{(\omega_L-\omega_{L'})^2}{2|k-k_{L'}|^2 {\mathrm v}_{Ti}^2}\right)} \times \notag
\\\left[\frac{1}{k_BT_i}\frac{\omega_{L'}-\omega_L}{\omega_L}W(k_{L'})W(k)\right]
\end{align}
where $k, k_{L'}, \omega_L, \omega_{L'}$ are the wavenumber and frequency of the initial and scattered Langmuir waves, and
\begin{equation}
\alpha_{ion}=\frac{\sqrt{2\pi} \omega_{pe}^2}{4 n_e {\mathrm v}_{Ti} (1+T_e/T_i)^2},
\end{equation} with ${\mathrm v}_{Ti}=\sqrt{k_B T_i/M_i}$ the ion thermal speed, and $M_i$ the mass of a plasma ion.
It is evident from the exponential factor that the scattering is strongest for $\omega_L \simeq \omega_{L'}$ and $k_{L'} \simeq -k$,
i.e. for backscattering of the waves. The resulting momentum change for the Langmuir wave is absorbed
by the ions which we assume to have a Maxwellian distribution at temperature $T_i$.
This momentum transfer is small, so the deviation of the ion distribution from thermal can be neglected \citep{1995lnlp.book.....T}.

The second source term describes Langmuir wave decay, and is given by
\begin{align}\label{eqn:4ql_sSrc}
&{\rm St}_{decay}(W(k))=\alpha_S\omega_{k} \int dk_S \omega_{k_S}^S \times \notag \\ &
\Biggl[ \left(
\frac{W(k_L)}{\omega^L_{k_L}}\frac{W_S(k_S)}{\omega^S_{k_S}}-
\frac{W(k)}{\omega^L_k}\left(\frac{W(k_L)}{\omega^L_{k_L}}+
\frac{W_S(k_S)}{\omega^S_{k_S}}\right)\right)\delta (\omega^L_{k}-\omega^L_{k_L}-\omega^S_{k_S})\Biggr.
\notag \\&
-\left.
\left(
\frac{W(k_{L'})}{\omega^L_{k_{L'}}}\frac{W_S(k_S)}{\omega^S_{k_S}}-
\frac{W(k)}{\omega^L_k}\left(\frac{W(k_{L'})}{\omega^L_{k_{L'}}}-
\frac{W_S(k_S)}{\omega^S_{k_s}}\right)\right)\right.\times \notag \\ & \Biggl.\delta(\omega^L_{k}-\omega^L_{k_{L'}}+\omega^S_{k_S}) \Biggr],
\end{align}
where $W_S(k_S), \omega_{k_S}^S$ are the spectral energy density and frequency of ion-sound waves, given by $\omega_{k_S}^S=k_S {\mathrm v}_s$ with ${\mathrm v}_s=\sqrt{k_BT_e(1+3T_i/T_e)/M_i}$ the sound speed, and the constant is
\begin{equation}
\alpha_S=\frac{\pi \omega^2_{pe}(1+3T_i/T_e)}{4n_ek_B T_e}.
\end{equation}

{For a given initial Langmuir wavenumber, $k$, we have two possible processes, namely $L \rightarrow L' +s$ and  $L +s \rightarrow L'$. The wavenumbers of the resulting Langmuir wave, $k_L, k_{L'}$ respectively, and the participating ion-sound wave, $k_S$, are found from simultaneous solution of the equations of energy conservation (encoded by the delta functions in Equation \ref{eqn:4ql_sSrc}), and momentum conservation, given by $k_L= k-k_S$ and $k_{L'}=k+k_S$ for the two processes respectively. For example, for the process $L \rightarrow L' +s$ we find $k_{L} \simeq - k$, and $k_S \simeq 2 k$, and the initial Langmuir wave is backscattered. More precisely, we have $k_{L} = - k + \Delta k$ with the small increment $\Delta k= 2 \sqrt{m_e/M_i} \sqrt{(1+3T_i/T_e)}/(3\lambda_{De})$. Thus repeated scatterings tend to accumulate Langmuir waves at small wavenumbers.}

\subsection{Ion-sound wave evolution}\label{sec:Ion}
The evolution of the ion-sound wave distribution is given by \citep[e.g.][]{1980MelroseBothVols, 1995lnlp.book.....T}
\begin{align}\label{eqn:4ql_s}&
\frac{\partial W_S(k)}{\partial t}=-\gamma_S(k)
W_S(k)\notag \\&-\alpha_S ({\omega^S_k})^2\int
\left(
\frac{W(k_L)}{\omega^L_{k_L}}\frac{W_S(k)}{\omega^S_{k}}-
\frac{W(k_{L'})}{\omega^L_{k_{L'}}}\left(\frac{W(k_L)}{\omega^L_{k_L}}+
\frac{W_S(k)}{\omega^S_{k}}\right)\right)\times\notag \\&
\delta(\omega^L_{k_{L\prime}}-\omega^L_{k_L}-\omega^S_k)dk_{L\prime}.
\end{align}
The second term here is analogous to Equation \ref{eqn:4ql_sSrc}, describing the {interaction of an ion-sound wave at wavenumber $k$ with a Langmuir wave at wavenumber $k_L$, producing a Langmuir wave at wavenumber $k_{L'}$.} {Again, these participating wavenumbers are found from simultaneous solution of energy (frequency) and momentum (wavenumber) conservation.} The first term is Landau damping of the waves,
with coefficient
\begin{equation}
\gamma_S(k)=\sqrt{\frac{\pi}{2}}\omega^S_k\left[\frac{{\mathrm v}_s}{{\mathrm v}_{Te}}+\left(\frac{{\mathrm v}_s}{{\mathrm v}_{Ti}}\right)^3\exp\left[- \left(\frac{{\mathrm v}_s^2}{2{\mathrm v}_{Ti}^2 }\right) \right]\right].
\end{equation}
Using the definitions of ${\mathrm v}_s, {\mathrm v}_{Te}, {\mathrm v}_{Ti}$, we see that the {ion contribution}  dominates
and $\gamma_s\simeq \omega_k^S (3 + T_e/T_i )^{3/2}\exp{(-[3 + T_e/T_i])}$,
which is of the order of $\omega_k^S$ unless $T_i\ll T_e$, {which is therefore a condition leading to high levels of ion-sound waves.}

\subsection{Electromagnetic emission}\label{sec:EM}

We describe the radio emission in terms of its brightness temperature, $T_T$, which is defined from the Rayleigh-Jeans law for the radiation intensity as function of frequency by $I(\nu)=2 \nu^2  k_B T_T/c^2$. We consider radiation at positive and negative wavenumber, the former propagating outwards from the Sun, along the direction of the beam propagation and the latter propagating backwards, and thus consider brightness temperature averaged over the corresponding hemisphere.

For thermal radiation, the definition of the brightness temperature gives $T_T=T_e$ the plasma temperature.  This radiation level is maintained by the thermal bremsstrahlung emission from the plasma particles, and by a corresponding damping. Kirchoff's law says that these must be related by $P(k) = \gamma_d T_e$ for $P$ the thermal emission rate, and $\gamma_d$ the damping, in order that the two balance to give the thermal radiation level. Thus, from the bremsstrahlung damping rate, \begin{equation}\label{eqn:gammaD}\gamma_{d}(k)=\gamma_{c}\frac{\omega_{pe}^2}{\omega(k)^2},\end{equation}  we can derive the thermal emission rate, or vice versa.


For emission described in terms of its brightness temperature we {therefore} have simply
\begin{align}\label{eqn:EMEvol}&\frac{dT_T(k_T)}{dt} = \gamma_d T_e - \gamma_{d} T_T(k_T)  -\frac{{\mathrm v}_g^T}{d}T_T(k_T)\notag\\& + \mathrm{St}_{fund}^{lts}(k_T)+\mathrm{St}_{fund}^{ion}(k_T)+ \mathrm{St}_{harm}^{ll't}(k_T)\end{align} where the first term is the thermal emission rate and the second term is the collisional damping. The third term, in which ${\mathrm v}_g^T$ is the group velocity for electromagnetic waves, given by ${\mathrm v}_g^T= c^2 k_T/\omega_T$, and $d$ is the source region size, describes the escape of radiation from the source region using a simple ``leaky box'' model. The waves propagate at velocity ${\mathrm v}_g^T$ and so over a time $dt$, we will lose a fraction given by $ {\mathrm v}_g^T dt/ d$ from the source region. Finally, we have the source terms, describing the production of radio emission from Langmuir waves{, either at the fundamental, near $\omega_{pe}$ or the harmonic near $2\omega_{pe}$}. These are derived in the following two subsections.

\subsection{Fundamental electromagnetic source terms}\label{sec:Fund}

The processes for emission at the fundamental are $L\rightleftarrows t \pm s$ where $L,s$ are Langmuir and ion-sound waves, and $t$ is an EM wave, and $L + i\rightleftarrows t + i'$, for $i,i'$ an initial and scattered plasma ion. The probability of both processes \citep[e.g.][]{1995lnlp.book.....T} is maximised for an EM wave with wavevector perpendicular to the initial Langmuir wave. Assuming the beam-generated Langmuir waves have some small angular spread {in wavenumber space}, covering a solid angle of $\Delta \Omega$, and further assuming that they are uniform within this spread, {fundamental emission will be produced approximately isotropically.} The larger the solid angle $\Delta \Omega$, the better this assumption becomes.

The source term entering Equation \ref{eqn:EMEvol} for fundamental emission by the processes $L +s \rightleftarrows t $ and $L\rightleftarrows t + s$ is {again based on the general expressions in \citep[e.g.][]{1980MelroseBothVols, 1995lnlp.book.....T}. Rewriting in terms of the hemisphere-averaged brightness temperature, {given by} \begin{equation}\label{eqn:tbW} T_T(k_T)=\frac{W_T(k_T)}{2 \pi k_B k_T^2}
\end{equation} {for $W_T(k_T)$ the EM wave spectral energy density,}
and using the assumptions described in the previous paragraph gives
\begin{align}\label{eqn:FundS}&\mathrm{St}_{fund}^{lts}(k_T)=\frac{\pi \omega_{pe}^4{\mathrm v}_s\left(1+\frac{3T_i}{T_e}\right)}{ 24 {\mathrm v}_{Te}^2n_eT_e} \times \int dk_L  \notag \\
&\left\{ \left[\frac{W_S(k_S)}{\omega_{k_S}^S} \frac{2\pi^2}{ k_B k_L^2\Delta\Omega} \frac{W_L({k}_L)}{\omega_{k_L}^L}-\frac{T_T({k}_T)}{\omega_{k_T}^T}\left( \frac{ W_S(k_S)}{\omega_{k_S}^S} + \frac{W_L(k_L)}{\omega_{k_L}^L}\right)\right]\right. \times \notag \\ &\delta(\omega_{k_L}^L +\omega_{k_S}^S-\omega_{k_T}^T) +  \notag\\\
&\left.\left[\frac{W_S(k_{S'})}{\omega_{k_{S'}}^S} \frac{2\pi^2}{k_B k_L^2\Delta\Omega} \frac{W_L({k}_L)}{\omega_{k_L}^L}-\frac{T_T({k}_T)}{\omega_{k_T}^T}\left( \frac{ W_S(k_{S'})}{\omega_{k_{S'}}^S} - \frac{W_L(k_L)}{\omega_{k_L}^L}\right)\right]\right.\times\notag \\&\left. \delta(\omega_{k_L}^L -\omega_{k_{S'}}^S-\omega_{k_T}^T)\right\}, \end{align}
where the participating wavenumbers are obtained from energy and momentum conservation. However, in this case the momentum conservation condition must be obtained from} the 3-D description of the process. {Using our assumption that the EM emission occurs approximately perpendicular to the initial Langmuir wave, the wavenumbers $k_L, k_S, k_T$ form a right-triangle and thus we find that $k_S^2=k_T^2 + k_L^2$.}


{Similarly, direct ion scattering, $L + i \rightleftarrows t +i'$, is described by
\begin{align}\label{eqn:FundIon}&\mathrm{St}_{fund}^{ion}(k_T)=\int {\mathrm d} k_L \frac{\sqrt{\pi} \omega_{pe}^2}{4 k_L n_e {\mathrm v}_{Ti} (1+T_e/T_i)^2} \exp{\left(-\frac{(\omega^T_{k_T}-\omega^L_{k_L})^2}{2k_L^2 {\mathrm v}_{Ti}^2}\right)}\times \notag\\&\left[\frac{\omega^T_{k_T}}{\omega^L_{k_L}}\frac{2\pi^2W_L(k_L)}{ k_B k_L^2}-T_T(k_T)-\frac{(2\pi)^3}{k_B T_i}\frac{\omega^T_{k_T}-\omega^L_{k_L}}{\omega^L_{k_L}}\frac{T_T(k_T)W_L(k_L)}{\Delta\Omega k_L^2}\right],
\end{align}
where again $\Delta \Omega$ is the angular spread of the Langmuir waves, and we have assumed again that the EM emission occurs approximately perpendicular to the initial Langmuir wave. The momentum change between initial and final waves is absorbed by the plasma ions, which are assumed to be thermal, as in the case of Langmuir wave scattering by ions considered above.}

\subsection{Harmonic electromagnetic emission source terms}\label{sec:Harm}

Emission at the harmonic of the plasma frequency occurs due to the coalescence of two Langmuir waves, $L + L' \rightleftarrows t$. Calculating an angle-averaged emission probability is complicated, because the emission probability depends directly on the participating wavenumbers, $k_1, k_2$ for the Langmuir waves, and $k_T$ for the EM wave, and the magnitudes of these depend on the geometry of the coalescence. In early models of emission, the ``head-on-approximation'' (HOA), where the Langmuir waves are almost antiparallel was suggested, which allows the emission probability to be greatly simplified. However, this assumption leads to significant overestimates of the emission rate, particularly at small wavenumbers \citep[e.g.][]{1979A&A....73..151M}, as it assumes that the   product electromagnetic wavenumber is far smaller than the initial Langmuir wavenumber, $k_T \ll k_L$, whereas in fact $k_T \simeq\left(\sqrt{3}{\mathrm v}_{Te}/c\right) k_{De}$ which is often comparable to $k_L$.



{Therefore the Langmuir waves cannot coalesce {exactly} head on, but rather at an angle less than $\pi$. Rather than specify this angle, which will depend on the values of $k_L$ and $k_T$, we specify the angle between one of the initial Langmuir waves and the final EM wave. The emission probability is quadrupolar \citep{1970nep..book.....T}, with a maximum when this angle is $\pi/4$, assuming both waves have the same sign for the beam-parallel-component. We therefore use this value to solve the wavenumber and frequency matching equations. We then average the probability over its FWHM using these wavenumber values. For Langmuir waves which are uniform over a solid angle (in wavenumber space) of $\Delta \Omega \gtrsim \pi/20$ and for $\omega_{EM} \gtrsim 2.01 \omega_{pe}$, the majority of Langmuir waves can participate in coalescence with the specified geometry, and we have an accurate approximation to the angle-averaged emission rate.
}


{As in the case of fundamental emission, we use the general expressions of \citep[e.g.][]{1980MelroseBothVols, 1995lnlp.book.....T}, convert to brightness temperature using Equation \ref{eqn:tbW}, and use our assumed emission geometry. The resulting source term for harmonic emission is
\begin{align}&\mathrm{St}_{harm}^{ll't}(k_T) =\omega_{k_T}^T \frac{\pi \omega_{pe}^2}{48 m_e n_e {\mathrm v}_{Te}^2} \int d k_1 \frac{(k_2^2-k_1^2)^2}{4 k_2^2} \times \notag
\\ &  \left[\frac{2\pi^2}{k_B k_2^2 \Delta\Omega}\frac{W(k_1)}{\omega_{k_1^L}}\frac{W(k_2)}{\omega_{k_2}^L}-\frac{T_T(k_T)}{\omega_{k_T}^T} \left(\frac{W(k_1)}{\omega_{k_1}^L}+\frac{W(k_2)}{\omega_{k_2}^L}\right)\right]\times \notag \\& \delta(\omega_{k_1}+\omega_{k_2}-\omega_{k_T}^T),
\end{align}
where $k_1, k_2$ are the wavenumbers of the forwards and backwards coalescing Langmuir waves respectively, $\omega_{k_1}, \omega_{k_2}$ the corresponding Langmuir wave frequencies, and $k_T, \omega^T_{k_T}$ the wavenumber and frequency of the EM wave.} From energy and momentum conservation, described by the delta function and by $\vec{k}_1+\vec{k}_2=\vec{k}_T$ respectively, we have the conditions \begin{equation}k_1\simeq\frac{1}{2}k_T \cos\left(\frac{\pi}{4}\right) +\frac{1}{2} \sqrt{4 \frac{\omega_{pe}(\omega^T_{k_T} -2\omega_{pe})}{3 {\mathrm v}_{Te}^2} + k_T^2 \left(\cos^2\left(\frac{\pi}{4}\right)-2\right)},\end{equation} considering only terms up to second order in $k_T$, and using that $\omega_T \simeq 2\omega_{pe}$; and $k_2^2=k_1^2 +k_T^2-2k_1k_T \cos(\pi/4)$ under the angular assumptions described previously.

\subsection{Propagation and absorption of radio emission}\label{sec:Prop}

\begin{figure*}
 \centering
\includegraphics[width=60mm]{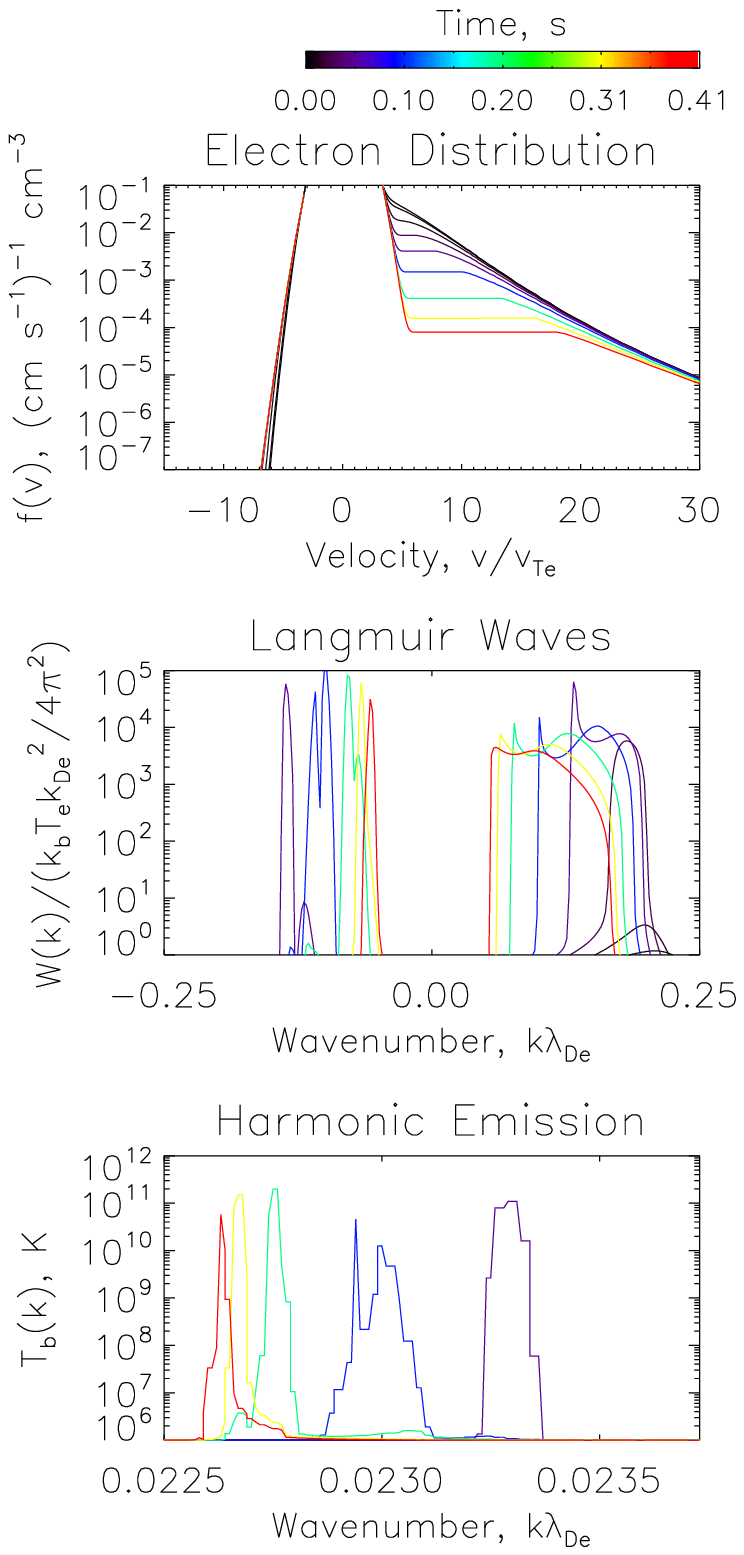} \includegraphics[width=60mm]{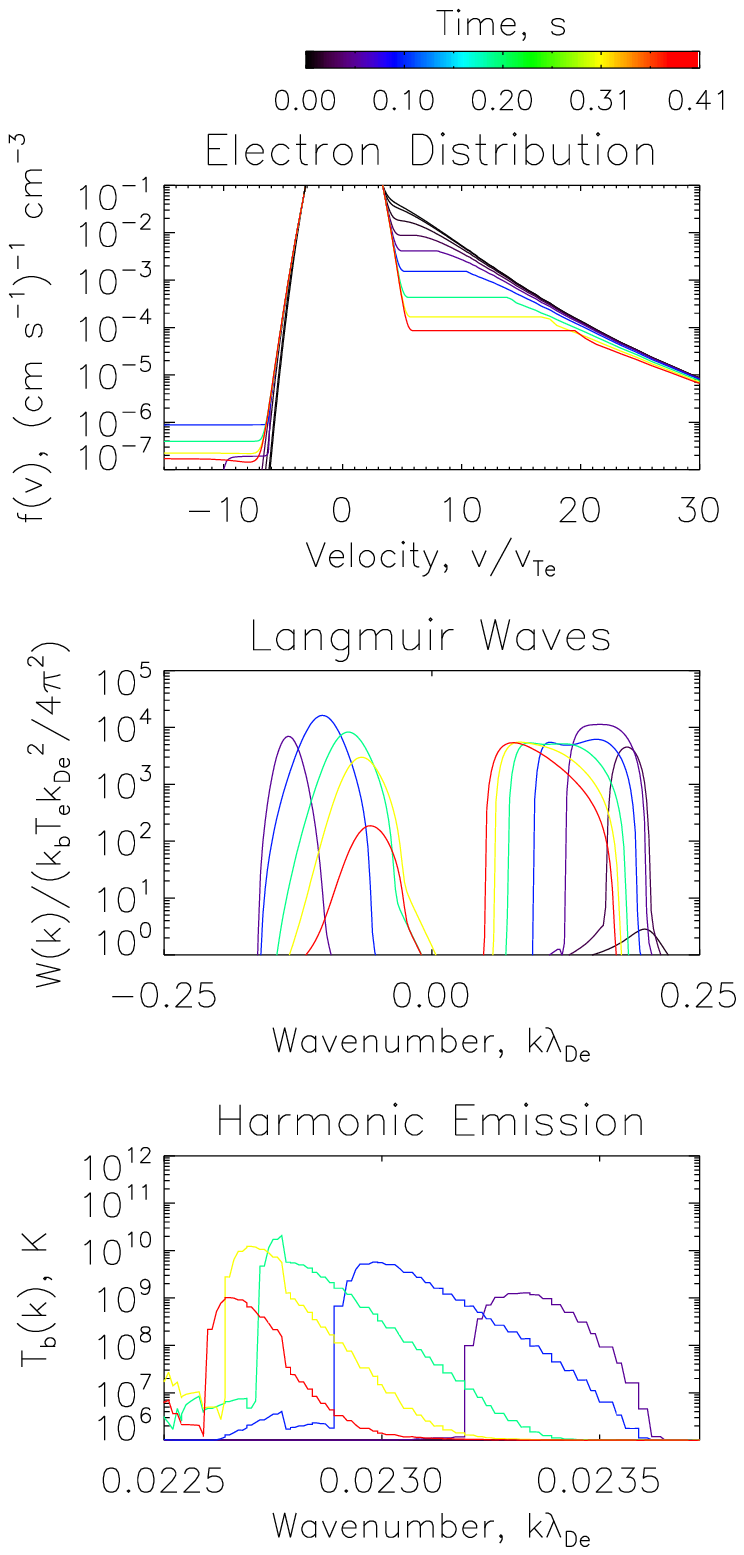} \\
\vspace{-0.6cm}
\includegraphics[width=60mm]{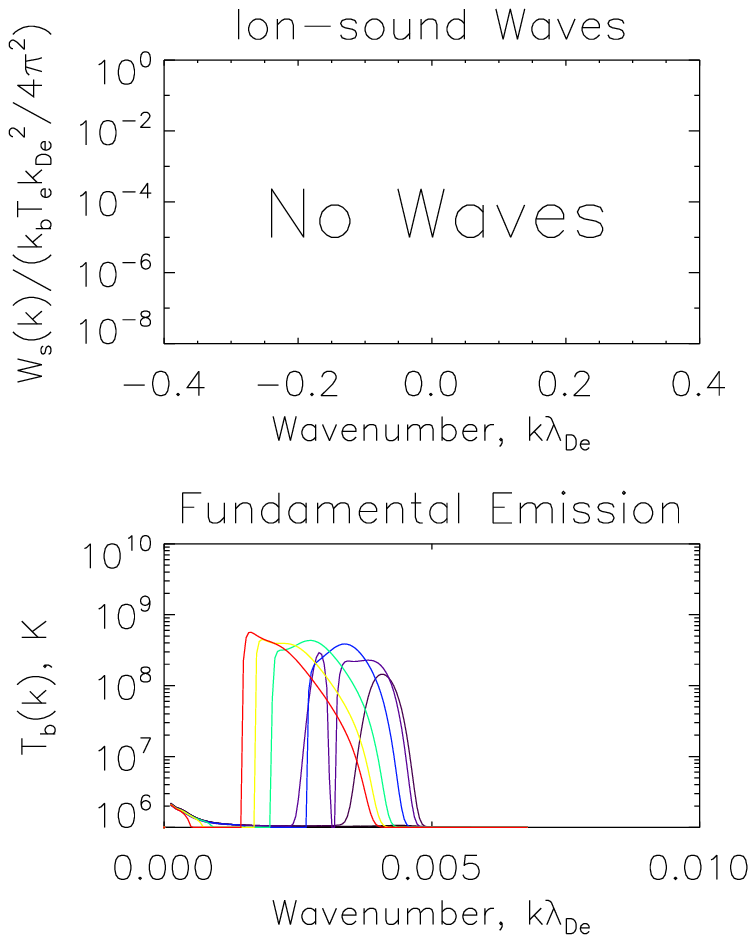} \includegraphics[width=60mm ]{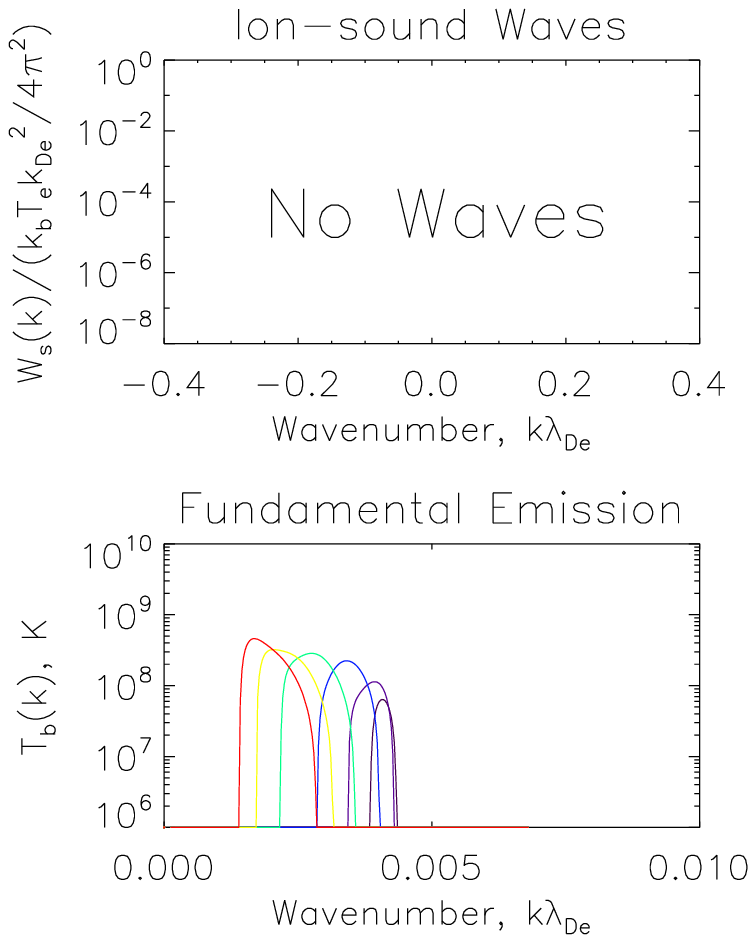}
\vspace{-0.6cm}
\caption{Top to bottom: the electron distribution function $f({\mathrm v})$; the spectral energy density of Langmuir waves $W(k)$; the {in-source} harmonic radio brightness temperature $T_T(k)$ {(for $k/k_{De}=0.0225$ to $0.0235$, corresponding to $\omega/\omega_{pe}=2$ to $2.08$)}; the spectral energy density of ion-sound waves $W_S(k)$; the {in-source} fundamental radio brightness temperature $T_T(k)$ {(for $k/k_{De}=0.$ to $0.01$, corresponding to $\omega/\omega_{pe}=1$ to $1.1$)} for a collisionally relaxing electron beam in {homogeneous plasma (left column) and in inhomogeneous plasma with $\tau_D\simeq 0.24$~s (right column)}. Each coloured line shows the distribution at a different time, as shown in the colour bar. }\label{fig:IShomO}
\end{figure*}

\begin{figure*}
 \centering
\includegraphics[width=0.47\textwidth]{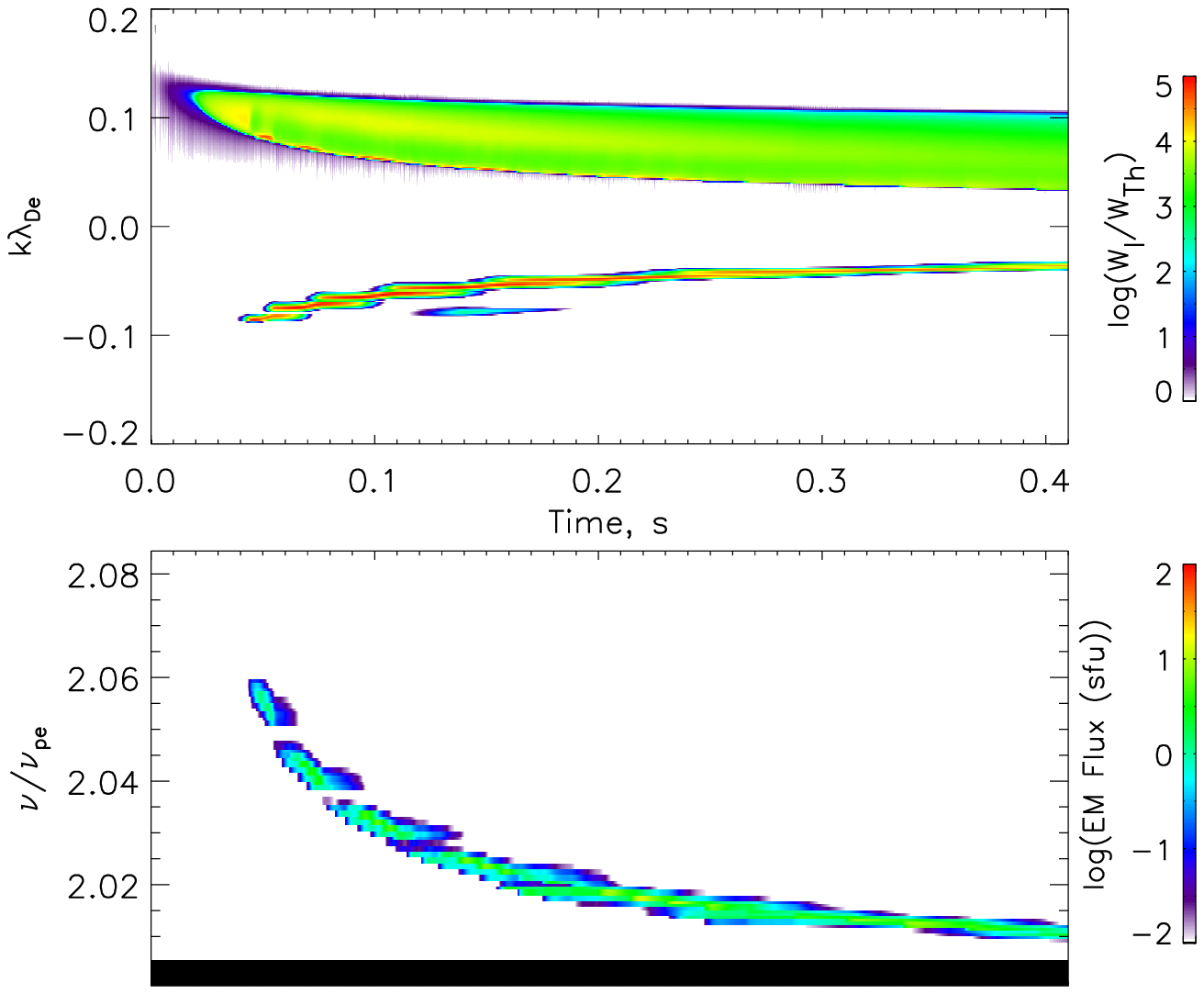}
\includegraphics[width=0.47\textwidth]{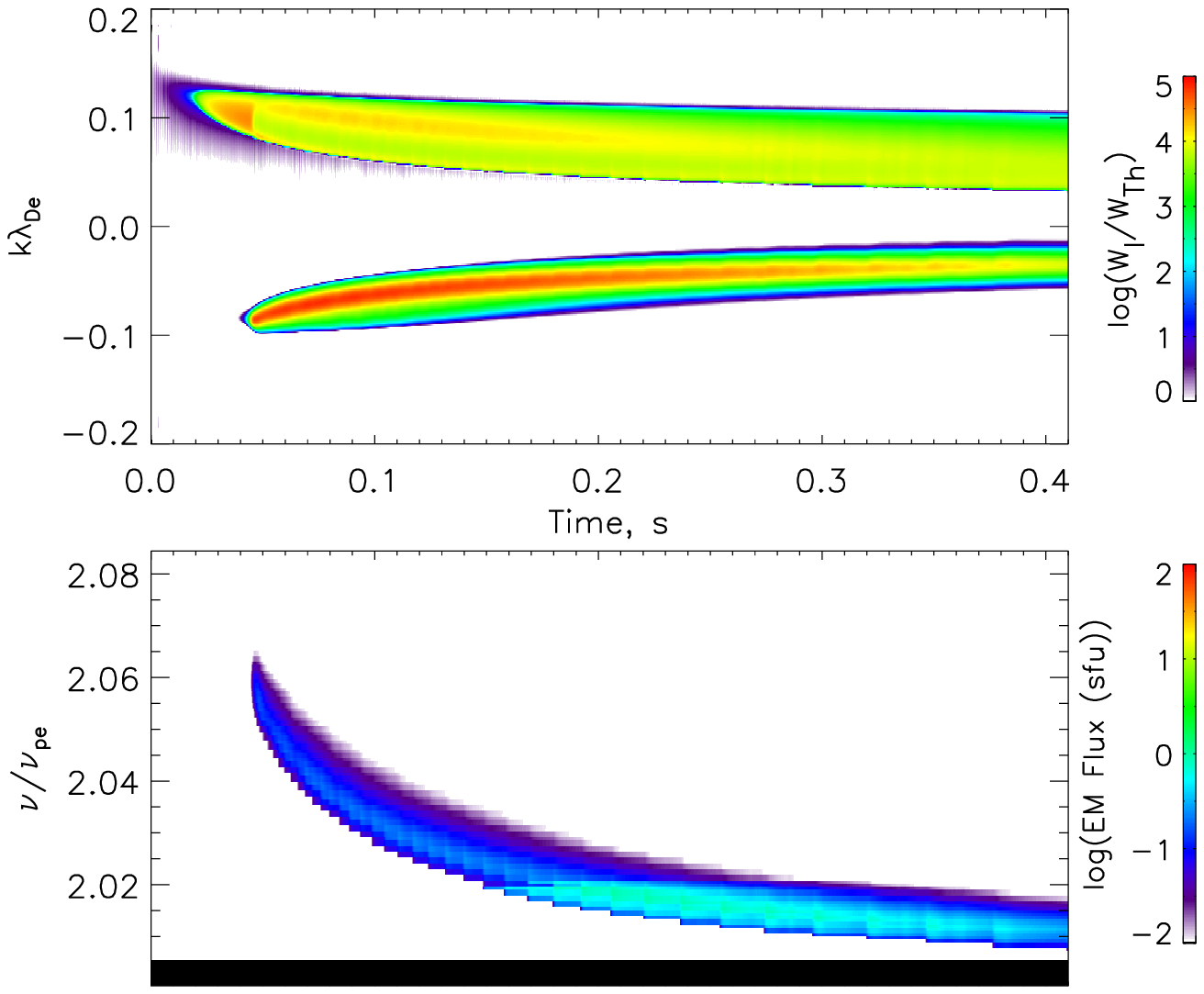}\vspace{0.3cm}\\
\includegraphics[width=0.47\textwidth]{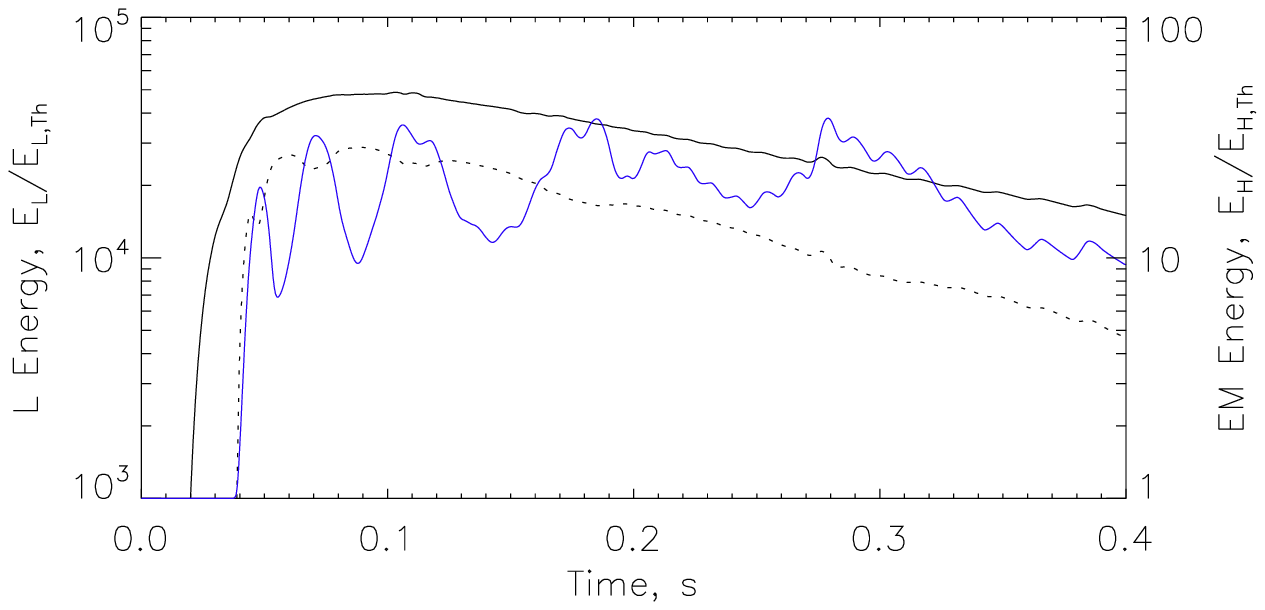}
\includegraphics[width=0.47\textwidth]{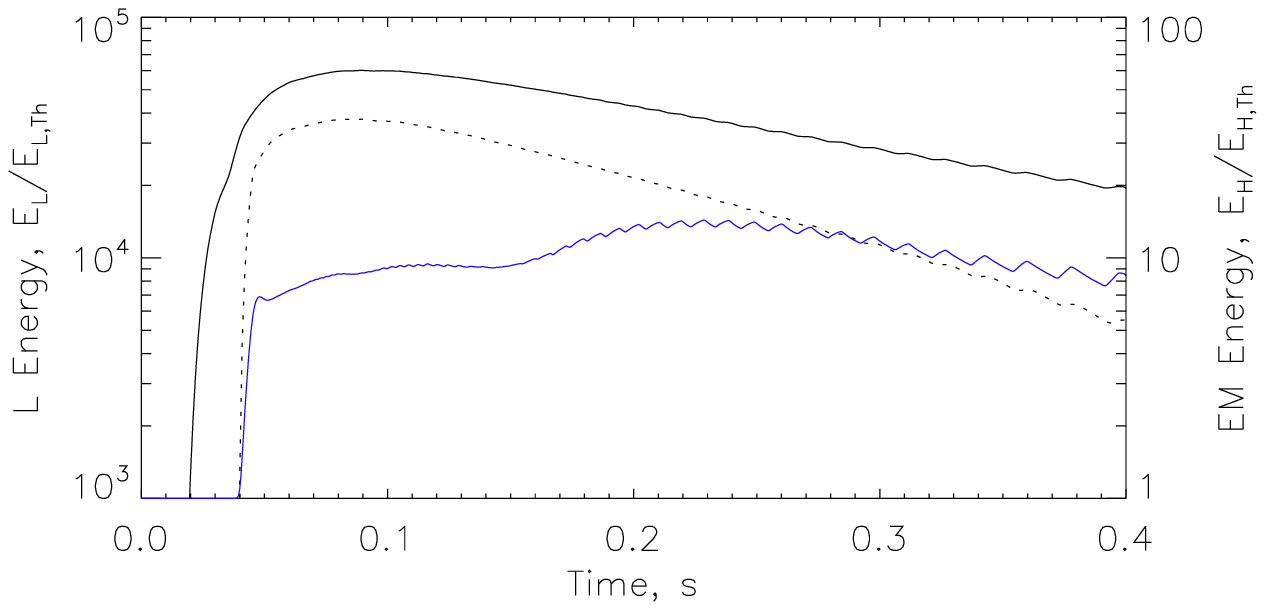}
\caption{The Langmuir wave spectral energy density and radio emission in sfu over the first 0.4~s of beam evolution. Top: Langmuir wave spectral energy density $W(k)$ normalised to the thermal level, against wavenumber $k$ on the vertical axis, and time on the horizontal axis. Middle: the radio flux in sfu as a function of frequency, $\nu=\omega/(2\pi)$, including the effects of absorption during propagation. The source size and plasma density profile are as described in the text, and the observed background flux from a thermal source of this size is $\sim 10^{-2}$~ sfu. Bottom: The total energy in Langmuir waves (solid line), backscattered Langmuir waves (negative wavenumber) only (dotted line), and radio emission (blue line), normalised by the thermal levels, against time. The left panel shows homogeneous plasma, the right, inhomogeneities with $\tau_D\simeq 2.4$~s.}\label{fig:IShom}
\end{figure*}

Finally, to relate the brightness temperature of radio emission within the source, given by the solution
of Equation \ref{eqn:EMEvol}, to the observed flux
we must consider the absorption of radiation during propagation.
Collisional absorption (inverse bremsstrahlung) with damping rate $\gamma_d$ gives an optical depth of
\begin{equation}\label{eqn:optDep}
\tau=\int_0^{1AU} \frac{\gamma_d(x)}{{\mathrm v}_g^T(x)} dx,
\end{equation}
where
${\mathrm v}_g^T=c/(1-\omega_{pe}^2/\omega^2)^{1/2}$ is the group velocity for electromagnetic waves.
Because this tends to zero as $\omega$ tends to $\omega_{pe}$, emission at the fundamental has a far larger
optical depth than that at the harmonic.

For emission at a frequency $\omega_0$ we then have
\begin{equation}\label{eqn:optDepOm0}
\tau(\omega_0)=\int_0^{1AU} \hspace{-0.5cm}dx  \sqrt{\frac{2}{\pi}}\frac{e^2 \ln\Lambda}{4 {\mathrm v}_{Te}^3 m_e c\omega_0}  \left(\frac{\omega_{pe}^4(x)}{\sqrt{(\omega_0^2-\omega_{pe}^2(x))}}\right).
\end{equation}
{We assume isothermal plasma between source and observer, with an exponential density
profile $n_e(x)=n_0 \exp(-x/H)$, where $x$ is the distance from the region of emission at density $n_0$, and $H$ is the density scale height.
The local plasma frequency is thus
\begin{equation} \omega_{pe}(x)=\omega_{pe}(0) \exp{\left(-\frac{x}{2H}\right)}.
\end{equation}
From Equation \ref{eqn:optDepOm0} we find \begin{align}\tau(\omega_0)=\int_0^{1AU} \hspace{-0.5cm}dx  \sqrt{\frac{2}{\pi}}\frac{e^2 \ln\Lambda}{4 {\mathrm v}_{Te}^3 m_e c\omega_0}  \left(\frac{\omega_{pe}^4(0) \exp{\left(-\frac{2x}{H}\right)}}{\sqrt{\omega_0^2-\omega_{pe}^2(0)\exp{\left(-\frac{x}{H}\right)}}}\right).
\end{align}
This may be integrated to find
\begin{align}\label{eqn:optDepFin}
&\tau(\omega_0)=\sqrt{\frac{2}{\pi}}\frac{e^2 \ln\Lambda}{3 {\mathrm v}_{Te}^3 m_e}\frac{H}{ c}\times \notag\\&\left[ \omega_0^2 - \frac{1}{\omega_0}\sqrt{(\omega_0^2-\omega_{pe}^2(0))}\left(\omega_0^2 +0.5\omega_{pe}^2(0) \right) \right],
\end{align}
where we have neglected the small term involving plasma frequency at 1AU.}

The resulting escape fraction $\exp(-\tau)$ is rather small for $\omega_{pe}/(2\pi)\sim 1$~GHz, and very dependent on the density scale height, $H$, chosen.
For the corona this is around $10^9$~cm which gives a fraction of around $1/50$ for harmonic emission,
and below $10^{-7}$ for the fundamental. A scale height of $6\times 10^8$~cm gives again a very small result for the fundamental,
and a value of $1/10$ for the harmonic. Alternately, if the emission comes from a dense loop embedded in less dense background plasma,
the escape fraction for both components may be increased \citep{1998SoPh..179..421K},
although that for the fundamental remains very small.

The source size is calculated by assuming a linear size of $10^9$~cm \citep[e.g.][]{2013ApJ...766...75J} at a distance of 1~AU,
which gives $0.2^\prime$. Then from the source brightness temperature, we may obtain the observed flux in sfu (Solar Flux Units, 1~sfu=$10^{-19}$erg s$^{-1}$cm$^{-2}$Hz$^{-1}$) using the definition of specific intensity, i.e. the Rayleigh-Jeans law, $I(\nu)=2 \nu^2  k_B T_T/c^2$, {where $\nu$ is the frequency in Hz,}
and that for the flux, $F(\nu)=I(\nu)\pi \theta^2$ where $\theta$ is the angular radius of the source, giving $\pi \theta^2$ the solid angle covered by it.
We assume the brightness temperature is constant throughout the source.
Thus the observed flux, including absorption during propagation, is given by
\begin{equation}\label{eqn:Flux}F(\nu)= 2 k_B T_T(\nu) \frac{\nu^2}{c^2} \pi \theta^2 \exp{(-\tau)}
\end{equation}
with the optical depth $\tau$ given by Equation \ref{eqn:optDepFin}.

\section{Numerical Results}

\subsection{Initial conditions}
We take a plasma density of $n_e \simeq 10^{10}$~cm$^{-3}$, corresponding to a local plasma frequency of $\nu_{pe}=\omega_{pe}/(2\pi)=1$~GHz,
and a plasma temperature of $T_e=1$~MK. The ion temperature $T_i$ is either $0.5 T_e$ or $T_e$ in the cases below.

The initial electron distribution is a power law smoothly joined to the Maxwellian core, with velocity ${\mathrm v}_b$ and a power law index of $\delta$ in energy space \begin{align}\label{eqn:initEl}
&f({{\mathrm v}}, t=0)= \frac{n_e}{\sqrt{2\pi} {{\mathrm v}}_{Te}} \exp\left(-\frac{{{\mathrm v}}^2}{2 {{\mathrm v}}_{Te}^2}\right) \notag\\&+\frac{2 n_{b}}{\sqrt{\pi}\, { {\mathrm v}}_b}\frac{\Gamma(\delta)}{ \Gamma(\delta-\frac{1}{2})}
\left[1+({ {\mathrm v}}/{ {\mathrm v}}_{b})^2\right]^{-\delta}
\end{align} where $\Gamma$ denotes the gamma function. We take a power law index of $\delta =4 $ \citep[e.g.][]{1985SoPh..100..465D} and a beam velocity of ${\mathrm v}_b=10 {\mathrm v}_{Te}$. {The low-energy turnover produced by the unit term in the square brackets prevents the divergence of the fast-electron component as ${\mathrm v}$ goes to zero. The effect on the velocities of interest for Langmuir wave generation are minimal.}

\begin{figure*}
 \centering
\includegraphics[width=0.47\textwidth]{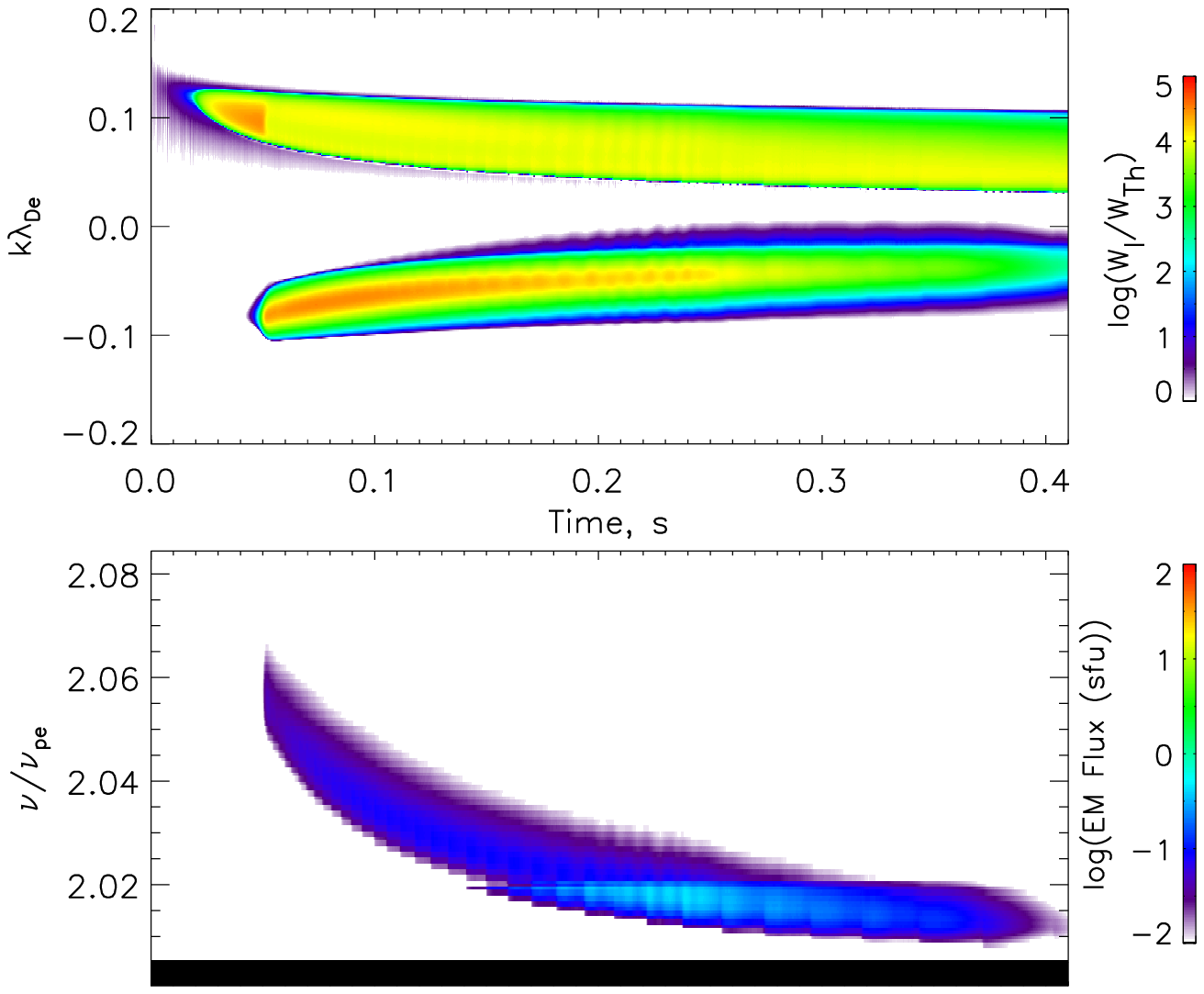}\includegraphics[width=0.47\textwidth]{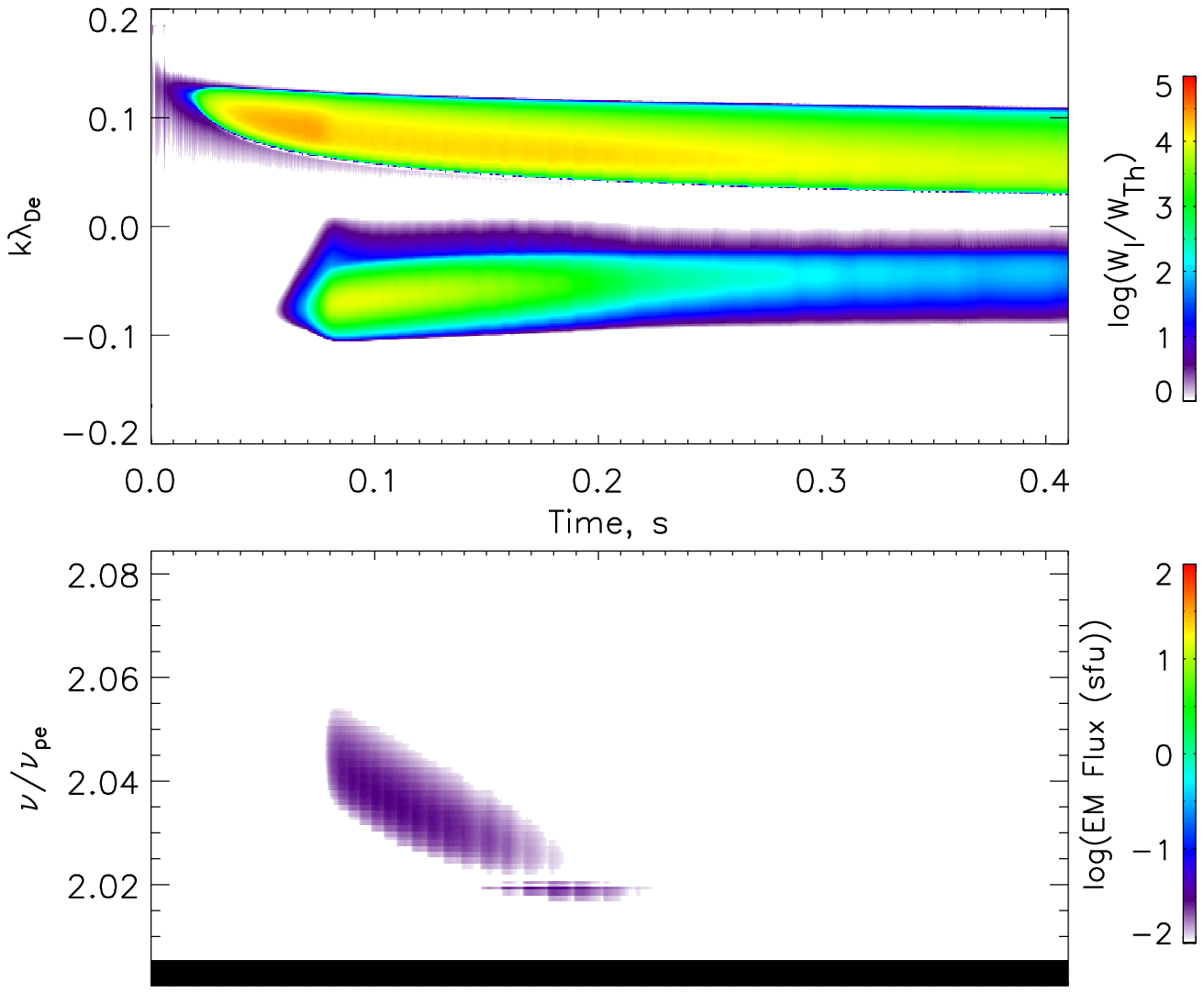}\vspace{0.3cm}\\
\includegraphics[width=0.47\textwidth]{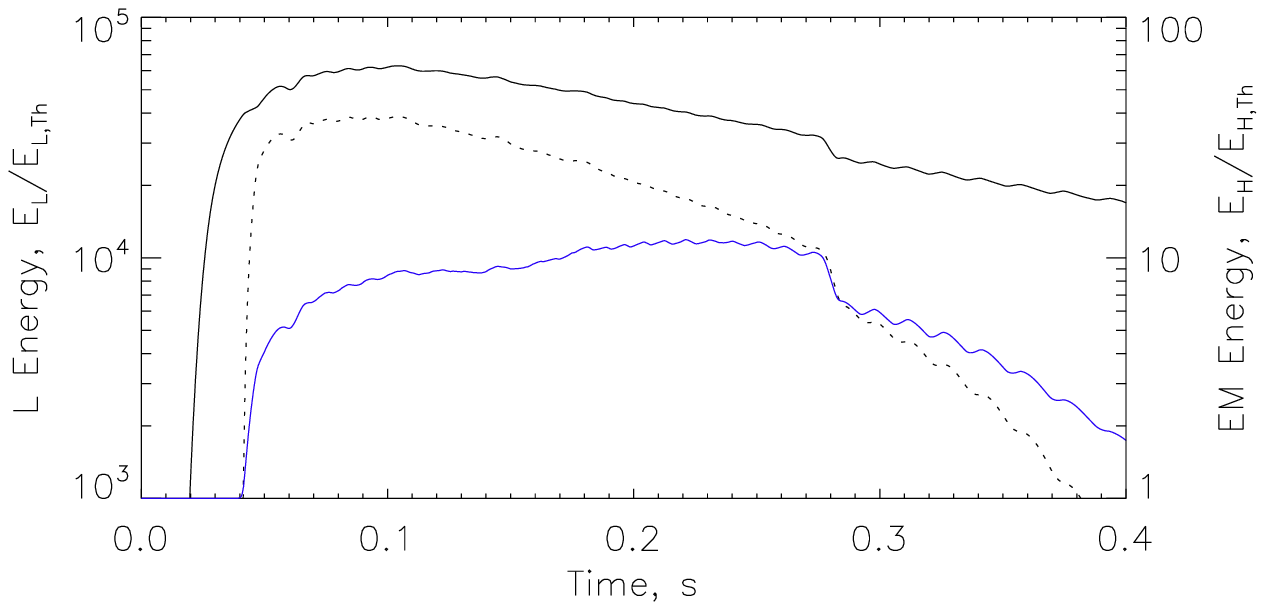}\includegraphics[width=0.47\textwidth]{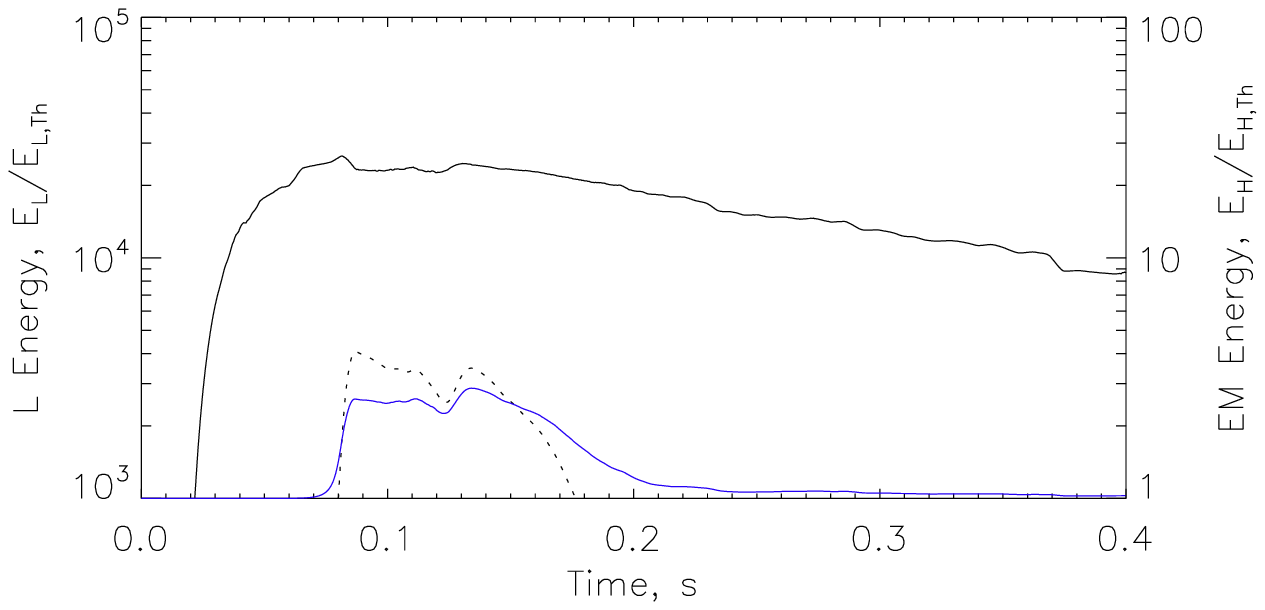}
\caption{As Figure \ref{fig:IShom} for inhomogeneous plasma with $\tau_D=0.24$~s (left) and $0.05$~s (right) respectively.}\label{fig:ISinhom}
\end{figure*}

The thermal Langmuir wave level is found by setting the LHS of Equation \ref{eqn:4ql2} to zero and balancing the thermal (first and third) terms on the RHS, giving \begin{align}\label{eqn:LTh}
W(k, t=0)&= \frac{k_B T_e}{4 \pi^2}\frac{k^2\ln\left(\frac{1}{k\lambda_{De}}\right)}
{1+\frac{\ln \Lambda}{16\pi n_e}\sqrt{\frac{2}{\pi}}k^3 \exp\left(\frac{1}{2k^2\lambda_{De}^2}\right)} \notag \\&\simeq \frac{k_B T_e}{4 \pi^2}k^2\ln\left(\frac{1}{k\lambda_{De}}\right),
\end{align} with $\lambda_{De}=1/k_{De}={\mathrm v}_{Te}/\omega_{pe}$. The initial level of ion-sound waves is thermal \citep{KRB}, \begin{equation} W_S(k_S)=k_B T_e k_{De}^2 \frac{k_{De}^2}{k_{De}^2+k_S^2}\end{equation} and the initial EM brightness temperature is thermal, $T_T=T_e$.
\begin{figure*}
 \centering
\includegraphics[width=0.27\textwidth]{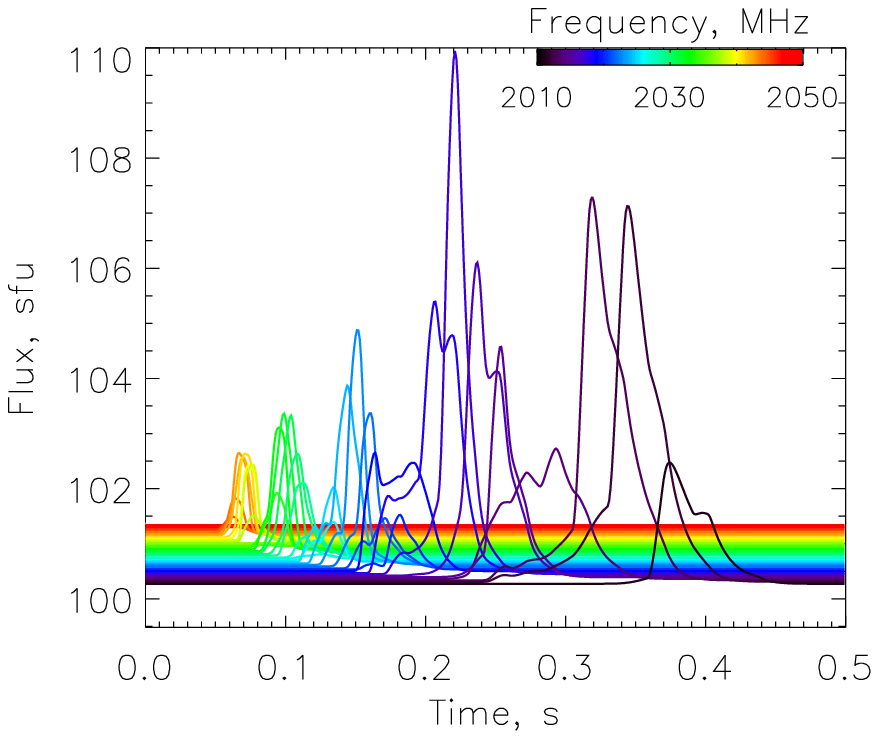}
\includegraphics[width=0.27\textwidth]{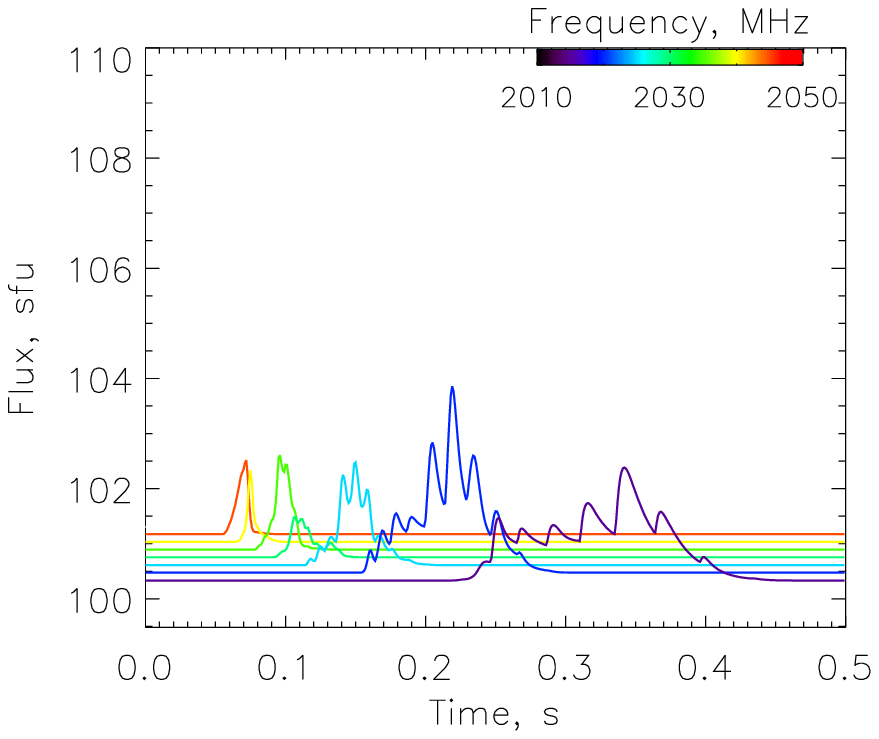}
\includegraphics[width=0.27\textwidth]{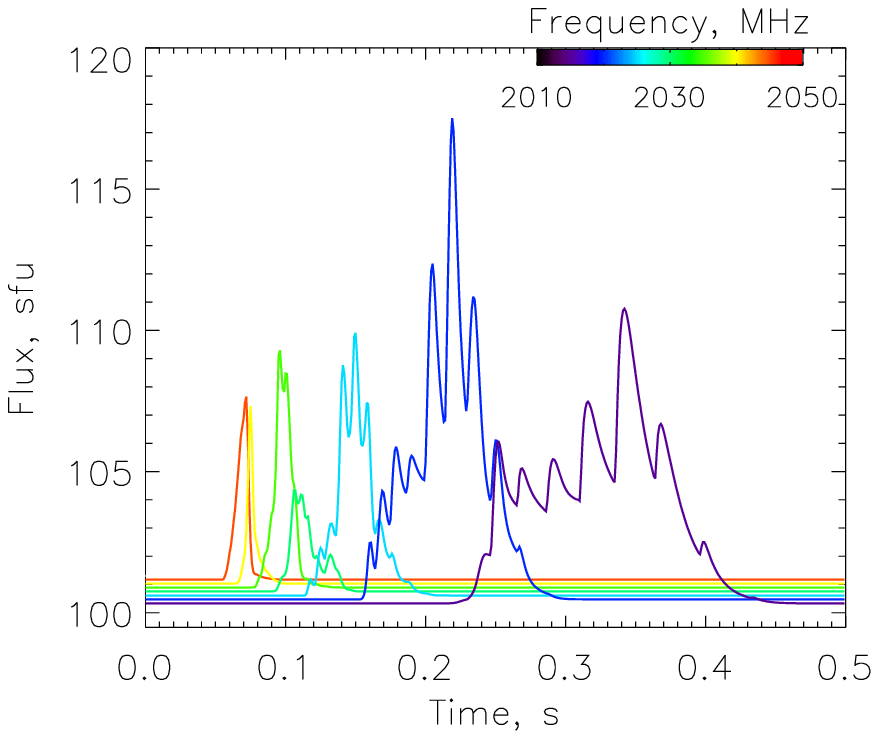}
\caption{Time profiles of the observed radio flux {(the fundamental component at 1~GHz is not visible due to strong absorption between source and observer)} for escape with coronal density scale height of $H=10^9$~cm, in sfu including the quiet-Sun background in homogeneous plasma {(no density fluctuations)}, averaged over frequency bands $\nu$ to $\nu+\Delta \nu$ for $\Delta \nu$=1~MHz (left) and 5~MHz (middle) at $\nu$ as shown in the colour bar. The right panel shows $\Delta \nu=5$~MHz for $H=6\times 10^8$~cm.}\label{fig:Spec}
\end{figure*}

\begin{figure*}
 \centering
\includegraphics[width=0.27\textwidth]{SpectroNoDiff_rerunBGband5}\includegraphics[width=0.27\textwidth]{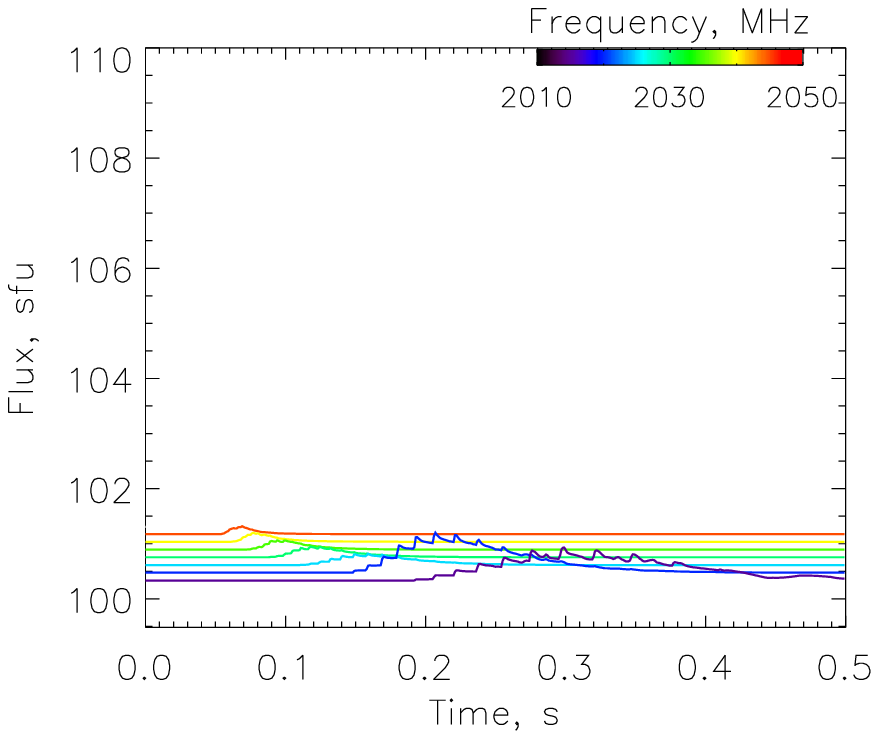} \includegraphics[width=0.27\textwidth]{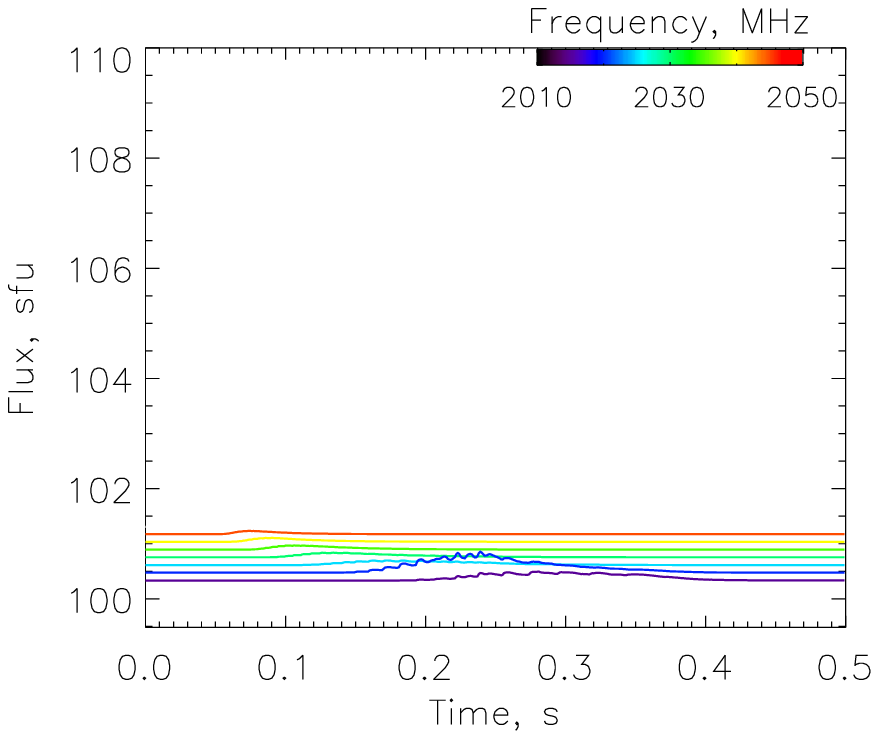} 
\caption{Time profiles of the observed radio flux in sfu including the quiet-Sun background, averaged over frequency bands $\nu$ to $\nu+\Delta \nu$ for $\Delta \nu=$ 5~MHz at $\nu$ as shown in the colour bar. Left to right, top to bottom:homogeneous plasma, and inhomogeneities with $\tau_D=2.4$~s and $0.24$~s. }\label{fig:SpecA}
\end{figure*}

\begin{figure}
 \centering
\includegraphics[width=0.4\textwidth]{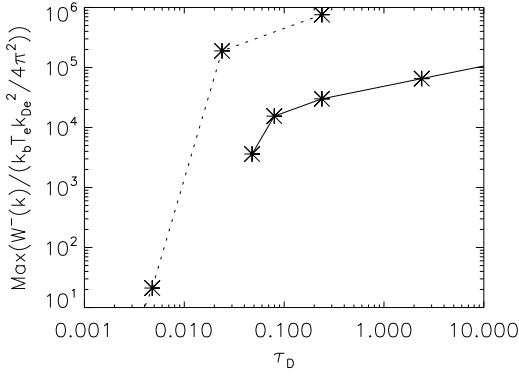}
\caption{The peak spectral energy density of backscattered Langmuir waves, $W^-(k)$,
against the timescales $\tau_D$ for several values of plasma inhomogeneity. The solid line shows plasma with equal ion and electron temperatures, while the dashed line has $T_i=0.5 T_e$.}\label{fig:peakBack}
\end{figure}

\begin{figure*}
\centering
\includegraphics[width=60mm]{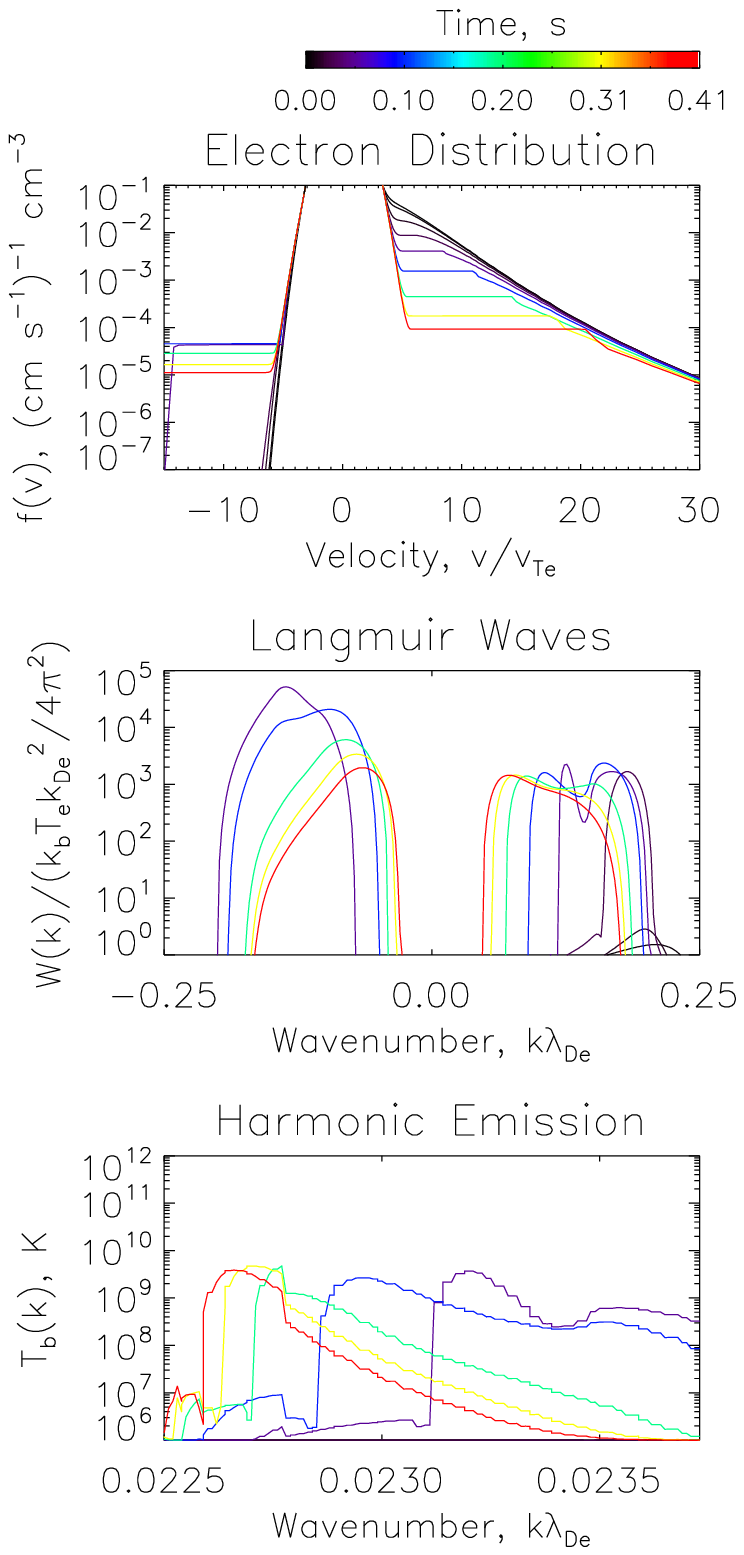} \includegraphics[width=60mm]{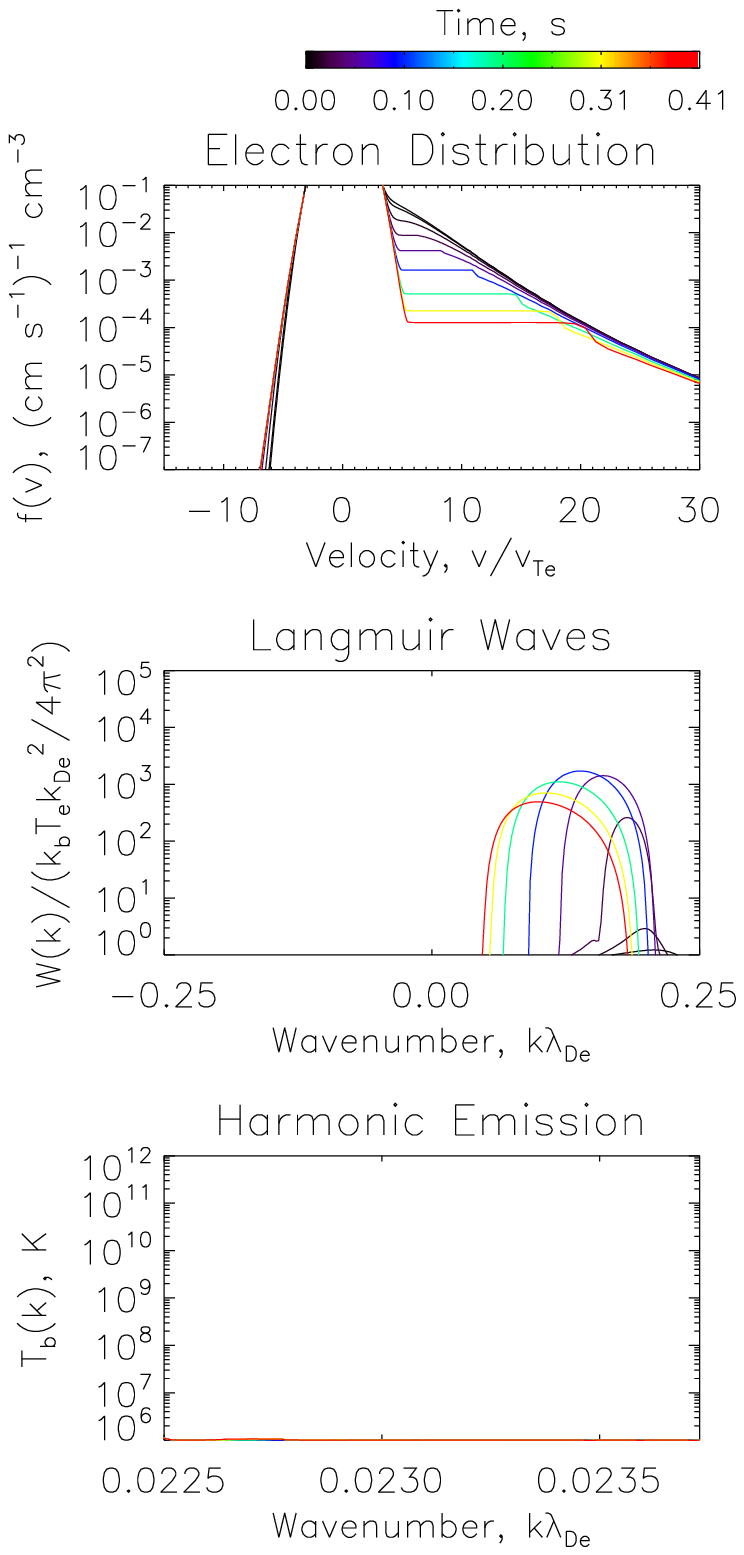}\\ 
\vspace{-0.6cm}
\includegraphics[width=60mm]{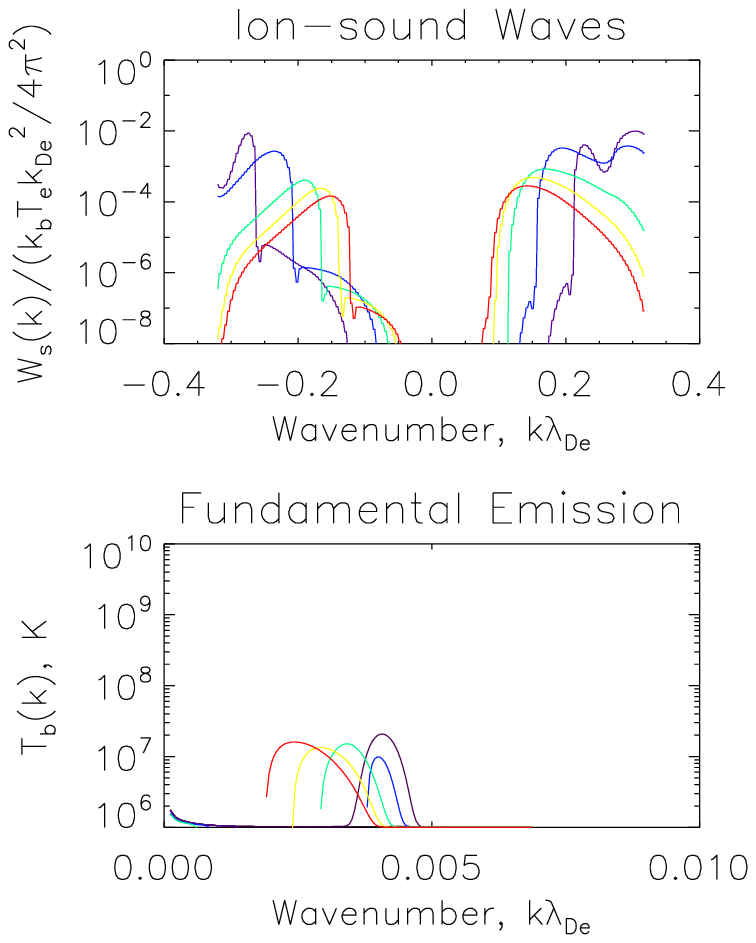} \includegraphics[width=60mm]{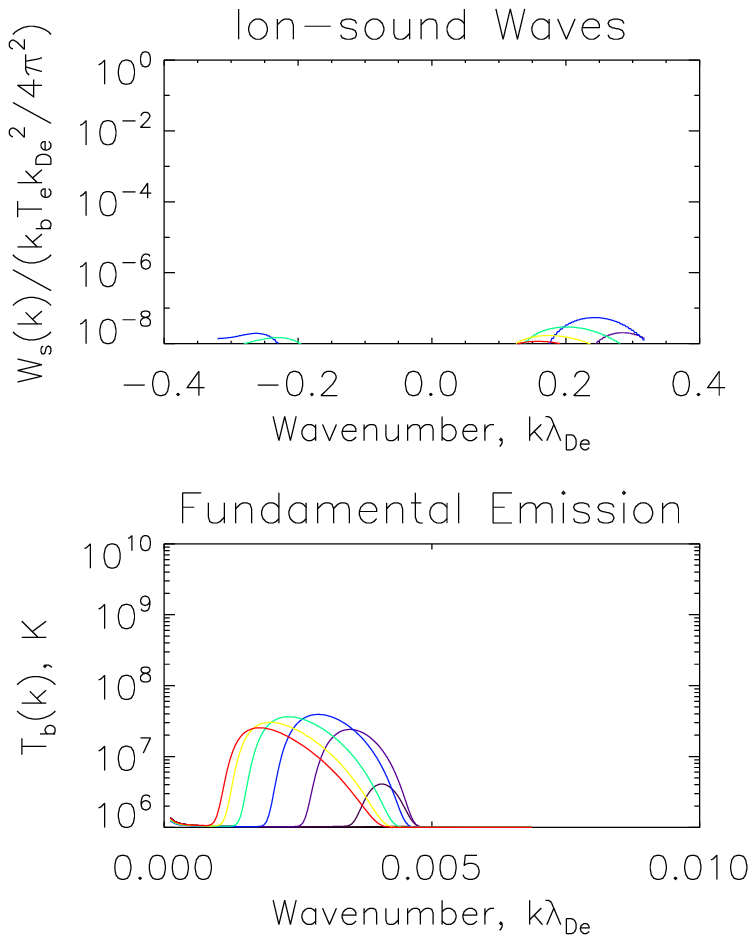}
 \caption{Top to bottom: the electron distribution function $f({\mathrm v})$; the spectral energy density of Langmuir waves $W(k)$; the {in-source} harmonic radio brightness temperature $T_b(k)$; the spectral energy density of ion-sound waves $W_S(k)$; the {in-source} fundamental radio brightness temperature $T_b(k)$ for a collisionally relaxing electron beam. Each coloured line shows the distribution at a different time, as shown in the colour bar. The left column shows the results in inhomogeneous plasma with $\tau_D\simeq 0.024$~s, the right with $\tau_D\simeq 0.0024$~s.}\label{fig:Snd0}
\end{figure*}


\subsection{Scattering by Ions}

In plasma with equal ion and electron temperatures, $T_i=T_e=1$~MK, ion-sound waves are strongly damped, and ion-scattering dominates the Langmuir wave backscattering. We take an initial beam as given by Equation \ref{eqn:initEl} with a beam density of $n_b=10^8$~cm$^{-3} \simeq 10^{-2} n_e$. {We note that this is for the initial power-law distribution. As collisional relaxation proceeds, a bump-on-tail distribution is produced, with a central velocity, density and width which vary over time due to continuing collisional losses. At this time, the majority of the beam electrons have thermalised, and so the level of Langmuir waves produced remains within the limits of weak-turbulence theory.}
In this section, we simulate electron-Langmuir wave evolution as given by Equations \ref{eqn:4ql1} and \ref{eqn:4ql2}, and radio emission as in Equation \ref{eqn:EMEvol}. {We omit the terms involving ion-sound waves (Equations \ref{eqn:4ql_sSrc} and \ref{eqn:FundS}) as the sound-wave damping rate is $\gamma_S \simeq \omega_S$, i.e the waves are damped over a timescale of less than one period, and the weak-turbulence equations will not accurately describe the behaviour \citep[e.g.][]{1980MelroseBothVols,1995lnlp.book.....T}. Instead we consider the process of scattering by plasma-ions only.}

Figure \ref{fig:IShomO} shows, for the case of homogeneous plasma, the resulting electron distribution, the Langmuir and ion-sound wave spectral energy densities, normalised to $k_B T_e k_{De}^2 /(4\pi^2)$ (see Equation \ref{eqn:LTh}), and the fundamental and harmonic radiation brightness temperatures. The fundamental emission reaches a brightness temperature of around $T_T\simeq 10^9$~K, while the harmonic emission can reach $T_T \simeq 10^{11}$~K. These are the in-source brightness temperatures, whereas the observable brightness temperature will be reduced by a factor of $\exp(-\tau)$, with $\tau$ the optical depth as in Equation \ref{eqn:optDepFin}. As noted above this means fundamental emission at these levels will not be observable, and the harmonic component will be reduced by around an order of magnitude. {The dispersion relation for electromagnetic waves is} $\omega=(\omega_{pe} + c^2 k^2)^{1/2}$ {and so the wavenumber variation shown here will lead to emission at a small range of
frequencies near $\omega_{pe}$ and $2\omega_{pe}$.}

Langmuir wave scattering by plasma ions is described by Equation \ref{eqn:LIonA}.
{By taking the limit of ${\mathrm v}_{Ti} \ll {\mathrm v}_{Te}$ in Equation \ref{eqn:LIonA}, we find the RHS is proportional to 
$d/dk (k W(k))$ \citep[as in][]{2003SoPh..215..335M}, and we therefore require locally positive gradient, $d/dk (k W(k))>0$ for the backscattered waves to grow.} Backscattering can then produce new regions of positive slope in the forwards wave spectrum, as seen in Figure \ref{fig:IShomO}, and these each produce a peak in the backscattered spectrum, leading to the multiple peaked distribution seen at around 0.1~s, and the similar structure seen in the harmonic emission.

{We now consider the effects of Langmuir wave diffusion in wavenumber due to density fluctuations, with coefficient $D(k)$ given by Equation \ref{eqn:diffcoeffGauss}. The parameters of small-scale density fluctuations are unknown in the low corona. However, previous work \citep{RBK} has shown that the most important factor for Langmuir wave evolution is the diffusive time-scale $\tau_D$, which we define here as}
\begin{equation}\label{eqn:tauD}
 \tau_D = \frac{k_{b}^2}{D(k_b)},
\end{equation} {where $k_b$ is the typical wavenumber of Langmuir waves in resonance with the reverse slope part of the electron distribution function. For a collisionally relaxing beam, this typical wavenumber varies over time, so we take the value at a time of $0.3 s$, $k_b \simeq 0.1 k_{De}$. We set the characteristic velocity to approximately the sound speed, ${\mathrm v}_0=10^7$~cm~s$^{-1}$, and consider $\tau_D\simeq 2.4$~s, $0.24$~s and $0.05$~s. }
To relate this $\tau_D$ to the RMS density fluctuation $\sqrt{\langle \tilde{n}^2\rangle}$ we must assume a value for $q_0$, the characteristic wavenumber of fluctuations. For example, taking $q_0=6 \times 10^{-4} k_{De}$ corresponding to a length scale of $10$~km, gives $\sqrt{\langle \tilde{n}^2\rangle} = 6 \times 10^{-4}$ for $\tau_D\simeq 0.24$~s. {This may be compared with observations in the higher corona and solar wind, which show levels of the order $\sqrt{\langle \tilde{n}^2\rangle} \sim 10^{-4}$ to $ 10^{-2}$ \citep{1972ApJ...171L.101C, 1979ApJ...233..998S}}.

Figure \ref{fig:IShomO} also shows the electron, Langmuir wave and ion-sound wave distributions and the fundamental and harmonic radio brightness temperature for $\tau_D \simeq 0.24$~s. Significant spreading of the Langmuir waves is evident, with a decrease in the backscattered wave level and consequently in the brightness of electromagnetic emission.

Figures \ref{fig:IShom} and \ref{fig:ISinhom} show the Langmuir wave spectral energy density and observable radio flux, calculated using Equation \ref{eqn:Flux}, in sfu (Solar Flux Units, 1~sfu=$10^{-19}$erg s$^{-1}$cm$^{-2}$Hz$^{-1}$) for three cases of density fluctuations, with $\tau_D=0.05, 0.24, 2.4$~s as well as the homogeneous case. The small oscillations in radio wave energy seen between 0.2 and 0.4 s are a numerical effect probably due to the discrete wavenumber grid used in the simulation code. The calculated fluxes are a few to perhaps ten sfu while the
duration of the simulated emission varies between approximately half a second for the homogeneous case down to around 0.1~s in the most inhomogeneous case. This duration is controlled by the Langmuir wave level, which in turn is controlled by the collisional relaxation of the electron beam.

Collisional relaxation also leads to the generation of Langmuir waves at smaller wavenumbers over time, which causes the emission to drift in frequency, in this case by approximately 50~MHz.
Electromagnetic emission is more efficient at lower frequencies, and so the level of emission rises. This is seen clearly in the bottom panels of Figures \ref{fig:IShom} and \ref{fig:ISinhom} where we plot the total energy in Langmuir waves, $E_L$, backwards (negative wavenumber) Langmuir waves $E_L^b$ and harmonic electromagnetic emission $E_H$, all normalised to the thermal levels. Initially, the rise of the electromagnetic wave energy $E_H$ closely tracks the backscattered wave energy $E_L^b$. However, for the two moderately inhomogeneous cases (top right and bottom left panels), we see the continued increase of electromagnetic wave energy after the backscattered Langmuir wave energy has peaked, between 0.15 to 0.3 seconds.

The observable thermal radio emission from our source, with size $0.2^\prime$ and temperature $10^6$~K, including absorption, is approximately $10^{-2}$~sfu. In contrast, the average whole Sun radio emission at 2~GHz varies over the solar cycle between approximately 50 and 150~sfu, and so we take a reference value of 100~sfu at 2~GHz, scaling with frequency as in \citet{2009LanB...4B..103B}. In Figure \ref{fig:Spec} we show the time profile of observed emission in homogeneous plasma including this background, averaged over frequency bands $\nu$ to $\nu+\Delta \nu$. Figure \ref{fig:Spec} shows the result in homogeneous plasma for $\Delta \nu$=1~MHz and 5~MHz and $\nu$ from 2.01~GHz to 2.05~GHz. The spiky nature of the emission in both time and frequency is evident. Figure \ref{fig:SpecA} shows the fluxes for $\Delta \nu$=5~MHz in homogeneous and inhomogeneous plasma. Inhomogeneity is seen to smooth out the time and frequency variations, as well as reducing the intensity of emission. The effects of absorption
during propagation are illustrated in Figure \ref{fig:Spec}, where we show density scale heights of $H=10^9$~cm and $6\times10^8$~cm, the latter giving a four-fold enhancement in the observed emission from the source.


\subsection{Ion-sound wave scattering}


For plasma with an electron temperature larger than the ion temperature, $T_e > T_i$ we can also consider the decay of Langmuir waves to an ion-sound wave, and a backscattered Langmuir wave, described by Equation \ref{eqn:4ql_sSrc}. We take $T_e=10^6$~K and $T_i=5\times 10^5$~K, with other parameters as in the previous section to obtain Figure \ref{fig:Snd0} which show the electron and wave distributions in inhomogeneous plasma with $\tau_D=0.024$~s and $\tau_D=0.0024$~s respectively. The level of backscattered Langmuir waves in these cases is far less affected by inhomogeneity than that due to ion-scattering alone, with complete suppression only for the shortest $\tau_D$. The harmonic radio emission we see is of similar magnitudes to that seen in the previous section, but covers a slightly wider range in $k$-space, and therefore frequency, at a given time. The fundamental emission is again much weaker than the harmonic.

\section{Discussion and conclusions}

The numerical results presented in Figures \ref{fig:IShomO} and \ref{fig:Snd0} show the tendency of plasma density inhomogeneities to suppress Langmuir wave backscattering.
In Figure \ref{fig:peakBack} we plot the peak backwards Langmuir wave spectral energy density for several cases
of density fluctuations, in both equal temperature plasma and plasma with $T_i=0.5 T_e$. Both show a sharp decrease
in backscattered level, but in the latter case this occurs only for much stronger inhomogeneity than the former.
This is due to the relative efficiencies of the two backscattering processes, or equivalently their timescales.
In general, the Langmuir wavenumber diffusion can suppress a process when $\tau_D$ is similar to {the timescale of the process},
as found in previous work \citep{RBK} in relation to the interaction of Langmuir waves with beam electrons.

In the cases shown here, the inhomogeneity is too weak to significantly suppress the beam-plasma interaction,
although we do see some slight electron acceleration, {due to the transfer of energy from slower to faster electrons by the wavenumber diffusion. Comparing Figure \ref{fig:Snd0} with the homogeneous case in Figure \ref{fig:IShomO}, the latter displays a broadened plateau
in the electron distribution around $20 \mathrm{v}_{Te}$, where electrons have been accelerated.}
Stronger inhomogeneities, with $\tau_D \sim \tau_{ql}$ the quasilinear time for beam-plasma interaction, can lead
to significant electron distribution tail self-acceleration.  Here, we have seen that plasma emission from
such a beam would be suppressed at {GHz frequencies}.

In-source brightness temperatures of $10^{11}$~K were seen in homogeneous plasma, corresponding to an observed flux of the order of a few sfu, {potentially} observable against the quiet Sun background. Even a low level of inhomogeneities can reduce these values by an order of magnitude, while strong fluctuations suppress the emission to the thermal level. The level of density fluctuations as commonly observed in the corona \citep{1972ApJ...171L.101C, 1979ApJ...233..998S}
with  $q_0^{-1}=10$~km leads to suppression of electromagnetic emission for RMS fluctuation magnitude of
around $\sqrt{\langle \tilde{n}^2\rangle} \sim 7\times 10^{-4}$ for equal temperature plasma,
and $\sqrt{\langle \tilde{n}^2\rangle} \sim 3\times 10^{-3}$ in plasma with larger electron than ion temperature.
Therefore we expect no observable plasma emission at these levels of inhomogeneity in the emission source.

To conclude, we have developed an angle-averaged emission model for fundamental
and harmonic plasma emission in collisional plasma, and {presented simulation results from this in homogeneous plasma and plasma with weak density fluctuations}. We find that the effects of density inhomogeneities on Langmuir wave backscattering can
be very significant even for low levels of plasma inhomogeneity. This has important effects not only on the Langmuir wave evolution itself,
but more significantly can suppress the production of plasma radio emission. For random density fluctuations with parameters as observed
in the solar corona, Langmuir wavenumber diffusion can completely suppress plasma radio emission even in the presence of strong
Langmuir wave turbulence. Moreover, these density fluctuations can simultaneously lead to electron self-acceleration,
increasing the number of fast electrons yet decreasing or completely hiding the often-expected
radio signature of Langmuir waves.

\begin{acknowledgements}

We thank the anonymous referee for the useful comments. 
This work was supported by an STFC STEP (Studentship Extension Programme) award (HR) and a STFC consolidated grant (EPK). Additionally, support by the Marie Curie PIRSES-GA-2011-295272 \textit{RadioSun} project, the European Research Council under the \textit{SeismoSun} Research Project No.~321141" is gratefully acknowledged.

\end{acknowledgements}

\bibliographystyle{aa}
\bibliography{refs1}

\end{document}